\documentclass{egpubl}
\usepackage{sca2026}

\SpecialIssuePaper         

\CGFccby

\usepackage[T1]{fontenc}
\usepackage{dfadobe}  

\usepackage{cite}  

\usepackage{amsmath,amssymb}
\usepackage{algorithm}
\usepackage{algorithmic}
\BibtexOrBiblatex
\electronicVersion
\PrintedOrElectronic
\ifpdf \usepackage[pdftex]{graphicx} \pdfcompresslevel=9
\else \usepackage[dvips]{graphicx} \fi

\usepackage{egweblnk} 

\title[Splitting Exact Coulomb Friction]%
      {A Splitting Architecture for Exact Reduced Coulomb Friction}

\author[H. Song, Y. Fan, U. Ascher \& D. Pai]
       {\parbox{\textwidth}{\centering
         Hongcheng Song\orcid{0009-0009-1784-4457},
         Ye Fan\orcid{0009-0005-0022-3364},
         Uri M. Ascher\orcid{0000-0002-5177-7181},
         and Dinesh K. Pai\orcid{0000-0002-5115-7230}}
        \\
        {\parbox{\textwidth}{\centering
          University of British Columbia, Canada}}
       }

\usepackage{xcolor}
\newif\ifrevmark\revmarkfalse
\newcommand{\revon}{\ifrevmark\color{red}\fi}
\newcommand{\revoff}{\ifrevmark\normalcolor\fi}
\usepackage{etoolbox}
\AtBeginEnvironment{table}{\ifrevmark\color{red}\fi}

\begin{document}


\maketitle
\begin{abstract}
Existing approaches to frictional contact dynamics typically either modify the Coulomb law to improve numerical robustness or solve the exact law in a fully coupled monolithic form. However, in its reduced form, exact Coulomb friction can be written as a cone complementarity problem with an augmented velocity, which reveals a natural split between a cone-constrained linear response and a scalar non-associated coupling induced by tangential velocity. We exploit this structure in the solver design. Our method uses an outer iteration to update the non-associated coupling explicitly, and an inner solve for a strongly convex cone-constrained quadratic program. This separation also makes the inner solver modular, so different numerical schemes can be used without changing the outer iteration. We evaluate the method on rigid-body benchmarks with stick-slip transitions and frictional stacking, and show that it reproduces exact Coulomb complementarity without smoothing or relaxing the friction law.


\end{abstract}
\section{Introduction}
\label{sec:introduction}

Coulomb friction gives a simple physical picture of contact: two bodies either separate, stick with the force inside the friction cone, or slide with the force on the cone boundary opposing tangential motion. After time discretization, this law can be written in reduced contact space as a complementarity relation between the contact reaction and an \emph{augmented contact velocity} introduced by 
De~Saxc\'{e} and Feng~\cite{de1998bipotential}. This compact form captures separation, stick, and slip in a single inclusion, and it is the exact reduced law used in nonsmooth mechanics when frictional accuracy matters. It is also the law behind familiar macroscopic behavior such as stable stacks, frictional support in arches, angles of repose that vary with the friction coefficient $\mu$, and clean stick--slip transitions.

One of the difficulties of the exact reduced law is the augmented velocity. Its normal component contains the magnitude of the tangential sliding velocity, so the law is \emph{non-associated}, implying that the frictional contact problem cannot be written as a single energy minimization. This scalar coupling entangles cone geometry to the rest of the dynamics, and any solver that treats the exact law as one monolithic object must carry that coupling throughout the solution process.

In computer graphics, there are generally three categories of approaches to handle this challenge. \emph{Smoothing} replaces the nonsmooth Coulomb law with a regularized force principle. This makes Newton-type iterations applicable, but it also introduces a stiffness accuracy trade-off which is determined by a smoothing scale that must be chosen in practice \cite{li2020incremental, geilinger2020add, larionov2024implicit}. \emph{Convex relaxation} drops the 
coupling of~\cite{de1998bipotential}  and solves the resulting cone-complementarity problem (CCP). This is convex and scales well, but it produces a visible artifact under sustained sliding, where contacts separate at the cone boundary even when the exact law would keep them in contact \cite{anitescu2002time, todorov2012mujoco, acary2018solving}. \emph{Monolithic nonsmooth Newton} keeps the exact law and tackles the full inclusion directly, solving a global semi-smooth problem whose machinery is harder to globalize, replace, or parallelize \cite{bertails2011nonsmooth, macklin2019non}.

The exact reduced 
law of~\cite{de1998bipotential} is standard in nonsmooth mechanics, but it is rarely the object that graphics solvers target explicitly. This law reveals a structural split. The reduced inclusion is the sum of two operators with very different numerical characters. The first is a cone-constrained linear response in the force. When the 
coupling is frozen, this part is a convex second-order cone QP. The second is the scalar map that carries the non-associated coupling. It is cheap to evaluate per contact but globally non-monotone. The point of our method is to treat these two pieces separately. It is a natural decomposition which has already been exposed by the law itself. Acary, Cadoux, Lemar\'{e}chal, and Malick~\cite{acary2011formulation} identified this decomposition algebraically in the mechanics literature. We adopt that and build a different solver around it.

We present a splitting architecture for the exact reduced Coulomb law based on this decomposition. The outer iteration follows Tseng's forward--backward--forward (FBF) template~\cite{tseng2000modified}. At each step, we explicitly evaluate the De~Saxc\'{e} and Feng coupling, and then, we solve a strongly convex cone QP with that coupling frozen and apply a lightweight correction that restores consistency with the exact law. This structure has two practical consequences. First, the inner cone QP is well-conditioned regardless of the rank of the Delassus operator, and it can be solved by whichever cone solver best matches the problem size and the available hardware. Second, the 
evaluation and correction of~\cite{de1998bipotential}
are parallel over contacts, and the inner solve can be warm-started across outer iterations. At convergence, the method satisfies the exact reduced law without introducing either a smoothing parameter or a convex relaxation.

We implement the method on top of NVIDIA Warp~\cite{warp2022} and the Newton physics engine~\cite{newton2025}, and evaluate it on rigid-body benchmarks including house-of-cards stacking, masonry arches under load, and turntable stick--slip. These examples are deliberately chosen to stress static support, frictional transmission, and mode transitions. In these regimes, smoothed and convex-relaxed solvers exhibit visible drift, while our splitting architecture preserves the exact reduced Coulomb law.

\paragraph*{Contributions.}
\begin{itemize}
\item A splitting solver architecture is presented for the exact reduced Coulomb friction law, which adopts the cone-QP / scalar-coupling decomposition of Acary et~al.~\cite{acary2011formulation} and replaces the Picard outer iteration of that work with a Tseng-style FBF scheme (\S\ref{sec:splitting}--\S\ref{sec:fbf}). This structure makes the inner cone solve modular, so, different convex SOCP solvers can be substituted without changing the outer loop.

\item A safeguarded adaptive step-size rule for the outer iteration (\S\ref{sec:stepsize}) is presented. Because the 
coupling of~\cite{de1998bipotential} is non-monotone, Tseng's original convergence theorem does not apply directly; hence, we combine a Lipschitz-based step size with a local-variation acceptance test and an asymmetric growth policy tuned to the stick-slip boundary.

\item A matrix-free, contact-parallel implementation is proposed (\S\ref{sec:implementation}) on top of NVIDIA Warp~\cite{warp2022} and Newton~\cite{newton2025}, together with an evaluation on rigid-body benchmarks (card-house, arch, turntable), which demonstrates reduced Coulomb behavior in stacking and stick-slip regimes where smoothed and convex-relaxed solvers visibly drift.
\end{itemize}

\section{Related Work}
\label{sec:related}

Frictional contact has been studied extensively in computer graphics~\cite{andrews2022contact}, robotics, and computational mechanics. Rather than survey this literature broadly, we focus on the methods most directly comparable to ours, the velocity-level time-stepping schemes for reduced contact-space dynamics with Coulomb friction. Within this scope, prior work differs mainly in how it handles the non-smooth and non-associated part of the law. Some methods smooth the law, some relax it to a convex cone-complementarity problem (CCP), and others solve the exact law monolithically. Our method follows a fourth route: it keeps the exact reduced law, but separates the cone-constrained linear response from the De~Saxc\'{e} coupling and treats them with different numerical tools.

\paragraph*{Smoothed and barrier-based friction.}
A large class of methods replaces Coulomb friction with a smooth surrogate so that each time step becomes an unconstrained or smoothly constrained minimization problem. Incremental Potential Contact~\cite{li2020incremental} and its rigid-body extension Rigid-IPC~\cite{ferguson2021intersection} combine barrier contact with a velocity-mollified and lagged friction force. ADD~\cite{geilinger2020add} smooth both the normal and tangential response so that the full dynamics is analytically differentiable. Larionov~et~al.~\cite{larionov2024implicit} solve smoothed friction fully implicitly with soft constraints and show that, in lagged-friction schemes, much of the inaccuracy comes from the lagging itself rather than from smoothing alone. These methods are attractive because the resulting objectives are smooth. They work well with Newton-type solvers, and integrate naturally with elastic dynamics and barrier-based non-intersection guarantees. The trade-off is that the solved law is a parametric approximation of Coulomb friction, with a smoothing parameter that mediates solve cost against frictional accuracy. Our method instead targets the unmodified reduced law.

\paragraph*{Convex relaxations and the CCP family.}
A second line of work keeps the cone geometry but removes the De~Saxc\'{e} coupling in order to recover a convex problem. In the formulation of Anitescu and Potra~\cite{anitescu2002time}, each time step is written as a cone complementarity problem (CCP). This matches Coulomb friction when the tangential velocity vanishes, but it no longer reproduces the exact law under sliding. The missing De~Saxc\'{e} term leads to the normal--tangential artifact often described as a boundary-layer effect~\cite{acary2018solving}. Despite this approximation, the CCP route is widely used because it is convex, scales to large contact sets well, and works well with mature first-order solvers. It underlies engines such as MuJoCo~\cite{todorov2012mujoco} and is commonly solved by projected fixed-point iterations~\cite{anitescu2010iterative} or, more recently, by ADMM with adaptive step scaling~\cite{tasora2021solving}. Our method uses a related cone-QP subproblem at the inner-solve level, but does not stop at the relaxed model. The outer correction reinstates the De~Saxc\'{e} and Feng coupling, so the converged iterate targets the exact reduced law rather than its convex approximation.

\paragraph*{Monolithic nonsmooth Newton.}
A third family of methods keeps the exact law and treats the resulting nonsmooth system as a single Newton-type problem. Bertails-Descoubes~et~al.~\cite{bertails2011nonsmooth} apply a generalized Newton method to the Alart--Curnier reformulation for hair assemblies. Macklin~et~al.~\cite{macklin2019non} extend nonsmooth Newton ideas to deformable multi-body dynamics and derive a symmetric Krylov-friendly system through a fixed-point reduction. Daviet~et~al.~\cite{daviet2011hybrid} combine an exact per-contact zero finder with an analytical fallback in hair simulation. More recently, Chen~et~al.~\cite{chen2024primal} bridge smooth and nonsmooth approaches through a logarithmic-barrier primal-dual interior-point formulation that approaches exact Coulomb friction as the barrier tightens. Most recently, Tsounis~et~al.~\cite{tsounis2025solving} benchmark a range of NCP solvers and implement a parallel algorithm in Kamino~\cite{kamino2026}, which we use as a baseline.  The main strength of this family is that it tackles the original inclusion directly and can exploit the fast local convergence of Newton iterations near a solution. Our method makes a different trade-off. Instead of solving the coupled nonsmooth system monolithically, we split it so that the implicit stage is always a strongly convex SOCP, leaving only the De~Saxc\'{e} and Feng coupling in the explicit part of the outer iteration.

\paragraph*{Splitting and fixed-point methods on the exact law.}
Closest to our work are methods that obey the exact law
of~\cite{de1998bipotential} and split the reduced problem into an inner cone solve and an outer update. Acary, Cadoux, Lemar\'{e}chal, and Malick~\cite{acary2011formulation} identify exactly this structure. The reduced inclusion can be written as a cone-constrained QP coupled to a scalar fixed-point map on the term of\cite{de1998bipotential}. They use this formulation to prove existence for the discrete problem and to derive a Picard outer iteration that freezes the coupling, solves the induced cone QP, and then updates the coupling at the new iterate. Our formulation~\eqref{eq:A}--\eqref{eq:B} is a similar decomposition, and we adopt it directly rather than introducing a new formulation of the law. Our contribution is on the solver side. We replace the Picard update with Tseng's FBF scheme~\cite{tseng2000modified}. In this setting, the correction after the cone solve is what makes the inner stage modular. The strongly convex cone QP can be handled by block Gauss--Seidel, projected gradient, ADMM~\cite{tasora2021solving}, or an interior-point solver such as Clarabel~\cite{Clarabel}, while the outer iteration itself remains the same.

Other two-level methods split frictional contact in different ways. The
staggered projections of Kaufman~et~al.~\cite{kaufman2008staggered} alternate a
normal-force QP with a tangential friction-disk projection until a joint fixed
point is reached; their split is therefore normal/tangential rather than
cone-QP/scalar-coupling. \revon Ly~et~al.~\cite{ly2020projective} incorporate dry
frictional contact into Projective Dynamics~\cite{pauly2014projective} by
splitting the PD global matrix, lagging the non-diagonal elastic coupling, and
using the resulting local relation to update nodal contact forces
semi-implicitly. This keeps the global PD solve unchanged, but the lagged
coupling is reconciled only through the local/global iteration. ADMM-style
solvers~\cite{overby2017admm,brown2018accurate} make a different split, using
auxiliary variables and dual updates to enforce consistency between the
subproblems. Our method also freezes one part of the contact law, but the
frozen term is the non-associated De~Saxc\'{e} coupling in reduced contact
space, while the contact response remains inside the cone QP. The FBF
correction~\eqref{eq:correction} then compensates for the change of this
coupling between the current and intermediate iterates, rather than leaving all
of the discrepancy to be removed by convergence of the outer loop.
Erleben~\cite{erleben2017rigid} formulates rigid-body contact in
proximal-operator form and derives PGS-style iterations from that viewpoint.\revoff
Daviet~\cite{daviet2020simple} develops a matrix-free Gauss--Seidel solver for
thin nodal objects within an ADMM splitting. Our matrix-free inner solver is
similar in spirit, but it appears inside an FBF outer correction for the
De~Saxc\'{e} term rather than as part of an ADMM dual update.

\paragraph*{Position of this work.}
Among methods that target the exact reduced Coulomb law, our contribution is a solver architecture built on the decomposition of Acary~et~al.~\cite{acary2011formulation}. We keep their cone-QP / scalar-coupling split, replace the Picard outer step with a Tseng-style FBF iteration, and use the resulting structure to make the inner cone solve modular, matrix-free, and parallel over contacts. The inner problem remains a standard strongly convex SOCP, so the solver used there can be chosen to match the contact count, sparsity pattern, and available hardware without changing the outer loop.
\section{Formulation}
\label{sec:formulation}
 
We consider the standard velocity-level time-stepping problem for rigid-body systems with frictional contact.
At time~$t$, the body positions and generalized velocities~$\mathbf{u}$ are known.
We seek the updated velocity~$\mathbf{u}^+$ at time $t{+}h$.
Given the mass matrix~$\mathbf{M}$, external forces~$\mathbf{f}$ evaluated at the known state (gravity), and the contact Jacobian~$\mathbf{J}$, the semi-implicit Euler step is
\begin{equation}\label{eq:timestep}
  \mathbf{M}(\mathbf{u}^+ - \mathbf{u}) = h\,\mathbf{f}
    + \mathbf{J}^\top \boldsymbol{\lambda}\,,
\end{equation}
\noindent where $\boldsymbol{\lambda}$ collects the 3-DOF contact reaction (two tangential, one normal) at each of the $n_c$ active contacts.
The unknowns are $\mathbf{u}^+$ and $\boldsymbol{\lambda}$, coupled by~\eqref{eq:timestep} and the Coulomb friction law described below.

\subsection{Reduced Contact-Space Problem}
\label{sec:reduced-kinematics}
 
Following standard practice~\cite{acary2008numerical, andrews2022contact}, we eliminate the primal velocities by substituting \eqref{eq:timestep} into the kinematic relation $\mathbf{v} = \mathbf{J}\,\mathbf{u}^+$, which gives the contact-space velocity at the new time as a function of the unknown reaction.
Defining the \emph{Delassus operator} $\mathbf{W} = \mathbf{J}\mathbf{M}^{-1}\mathbf{J}^\top$
and the \emph{free velocity} $\mathbf{v}_f = \mathbf{J}(\mathbf{u} + h\,\mathbf{M}^{-1}\mathbf{f})$, both computable from known data, the contact-space velocity becomes an affine function of the reaction:
\begin{equation}\label{eq:reduced}
  \mathbf{v}(\boldsymbol{\lambda}) = \mathbf{W}\boldsymbol{\lambda} + \mathbf{v}_f\,.
\end{equation}

This reduction is the exact contact-space image of the primal time step. The Delassus operator~$\mathbf{W}$ is symmetric positive semi-definite, and its application is matrix-free: each evaluation of $\mathbf{W}\boldsymbol{\lambda}$ requires one scatter ($\mathbf{J}^\top$), one diagonal solve ($\mathbf{M}^{-1}$), and one gather ($\mathbf{J}$), all parallelizable over contacts and
bodies.
Once $\boldsymbol{\lambda}$ is found, the body velocities are recovered by back-substitution into~\eqref{eq:timestep} and positions are advanced by semi-implicit Euler integration.

\subsection{Coulomb Friction and the De Saxc\'{e} and Feng Form}
\label{sec:desaxce-law}
 
The Coulomb friction law governs how contact reactions relate to contact velocities. For each contact $i$ with friction coefficient~$\mu_i$, three modes are possible:
\begin{equation}\label{eq:modes}
  m_i = \left\{
  \begin{array}{@{}l@{\qquad}l@{\quad}l@{}}
    \text{separation,} & v_{N,i} > 0, & \boldsymbol{\lambda}_i = \mathbf{0}\,,\\[4pt]
    \text{sticking,}   & \|\mathbf{v}_{T,i}\| = 0, & \boldsymbol{\lambda}_i \in \mathrm{int}\,K_{\mu_i}\,,\\[4pt]
    \text{sliding,}    & \|\mathbf{v}_{T,i}\| > 0, & \boldsymbol{\lambda}_i \in \partial K_{\mu_i}\,.
  \end{array}
  \right.
\end{equation}

\noindent Here $K_\mu$ is the second-order Coulomb friction cone, where the subscripts $N$ and $T$ denote the normal and tangential components, respectively:
\begin{equation}
    K_\mu = \{\boldsymbol{\lambda} :
\lambda_N \ge 0,\;\|\boldsymbol{\lambda}_T\| \le \mu\lambda_N\}
\end{equation}
 \revon \noindent Its interior $\mathrm{int}\,K_\mu = \{\boldsymbol{\lambda} : \lambda_N > 0,\;\|\boldsymbol{\lambda}_T\| < \mu\lambda_N\}$ collects the reactions that lie strictly inside the cone, where the tangential force stays below the Coulomb limit and the contact sticks. Its boundary $\partial K_\mu = \{\boldsymbol{\lambda} : \lambda_N \ge 0,\;\|\boldsymbol{\lambda}_T\| = \mu\lambda_N\}$ is the conical surface on which the friction force is fully mobilized and the contact slides.\revoff

These three cases can be stated compactly using the
\emph{augmented velocity} introduced by~\cite{de1998bipotential}:
\begin{equation}\label{eq:augmented}
  \tilde{\mathbf{v}}_i
    = \mathbf{v}_i + \mu_i\,\|\mathbf{v}_{T,i}\|\,\mathbf{e}_N\,,
\end{equation}

\noindent where $\mathbf{e}_N$ is the contact-normal unit vector.
With this definition, the exact Coulomb law for all contacts simultaneously reduces to the cone complementarity problem:
\begin{equation}\label{eq:soccp}
  K^*_\mu \ni \tilde{\mathbf{v}}(\boldsymbol{\lambda})
  \;\perp\;
  \boldsymbol{\lambda} \in K_\mu\,,
\end{equation}

\noindent meaning that $\tilde{\mathbf{v}}$ lies in the dual cone $K^*_\mu$, $\boldsymbol{\lambda}$ lies in the primal cone $K_\mu$, and $\tilde{\mathbf{v}}^\top\boldsymbol{\lambda} = 0$. This is equivalent to the standard disjunctive Signorini--Coulomb model~\cite{acary2008numerical}, but expressed as a single inclusion rather than a case-by-case enumeration.
 
Substituting~\eqref{eq:reduced} and~\eqref{eq:augmented} into~\eqref{eq:soccp} yields the \emph{exact reduced problem} that our solver targets:
\begin{equation}\label{eq:target}
  \mathbf{0} \in \mathbf{W}\boldsymbol{\lambda} + \mathbf{v}_f
    + \mu\,\|\mathbf{v}_T(\boldsymbol{\lambda})\|\,\mathbf{e}_N
    + N_{K_\mu}(\boldsymbol{\lambda})\,,
\end{equation}
\noindent where $N_{K_\mu}$ denotes the normal-cone operator of~$K_\mu$. The inclusion $-\tilde{\mathbf{v}}\in N_{K_\mu}(\boldsymbol{\lambda})$ is a restatement of the cone complementarity~\eqref{eq:soccp}, so the two forms describe the same condition. The operator $N_{K_\mu}(\boldsymbol{\lambda})$ is nonetheless distinct from the dual cone $K^*_\mu$: it depends on the point $\boldsymbol{\lambda}$, and is related to the dual cone by $N_{K_\mu}(\boldsymbol{\lambda}) = -K^*_\mu \cap \{\boldsymbol{\lambda}\}^{\perp}$.

\subsection{Source of the Difficulty}
\label{sec:difficulty}
 
The difficult term in~\eqref{eq:target} is the 
augmented velocity $\mu\|\mathbf{v}_T(\boldsymbol{\lambda})\|\,\mathbf{e}_N$:
it couples the tangential sliding velocity into the normal direction. Physically, this reflects the fact that Coulomb friction is \emph{non-associated}---the dissipation cannot be expressed through a single scalar potential~\cite{de1998bipotential}.
In optimization terms, this means that the frictional contact problem \emph{cannot be cast as a single energy minimization} over the friction cone.
 
Existing approaches handle this difficulty in one of three ways. \emph{Smooth approximation} methods replace the non-smooth Coulomb law with a smooth or barrier-based surrogate, changing the mathematical object being
solved~\cite{li2020incremental, geilinger2020add, larionov2024implicit}. \emph{Monolithic nonsmooth} methods keep the exact law but solve it as a single nonsmooth system, which can be difficult to globalize and
parallelize~\cite{bertails2011nonsmooth, macklin2019non}. \emph{Relaxation} methods drop the De~Saxc\'{e} and Feng term entirely, yielding a convex cone program (the so-called CCP formulation~\cite{anitescu2002time, todorov2012mujoco, tasora2021solving}) that is easier to solve but introduces artificial normal separation under sliding~\cite{acary2018solving, tasora2021solving}.
 
We take a different route: we \emph{keep the exact law but split it} so that the nonlinear coupling is handled explicitly while the cone geometry is handled implicitly. This is possible because~\eqref{eq:target} has a natural two-part structure---a cone-constrained linear response plus a scalar non-associated coupling---and this structure directly suggests the solver architecture described next.

\section{A Splitting Algorithm for Exact Reduced Coulomb Friction}
\label{sec:algorithm}
 
\subsection{Operator Splitting}
\label{sec:splitting}
 
The inclusion~\eqref{eq:target} can be decomposed as
$\mathbf{0} \in A(\boldsymbol{\lambda}) + B(\boldsymbol{\lambda})$,
with
\begin{align}
  A(\boldsymbol{\lambda})
    &:= \mathbf{W}\boldsymbol{\lambda} + \mathbf{v}_f
         + N_{K_\mu}(\boldsymbol{\lambda})\,,
    \label{eq:A}\\
  B(\boldsymbol{\lambda})
    &:= \mu\,\|\mathbf{v}_T(\boldsymbol{\lambda})\|\,\mathbf{e}_N\,.
    \label{eq:B}
\end{align}

\begin{itemize}
\item
$A$ contains the linear contact response with the cone constraint. When $B$ is frozen, $A(\boldsymbol{\lambda})=\mathbf{0}$ is precisely a cone-constrained quadratic program, the same mathematical object solved by CCP-based engines such as MuJoCo~\cite{todorov2012mujoco}.
 
\item
$B$ is the non-associated De~Saxc\'{e} and Feng coupling.
It is single-valued, cheap to evaluate per contact, and Lipschitz continuous with constant $L_B \le \mu_{\max}\,\|\mathbf{W}\|$.
It is, however, not monotone, which is a direct consequence of the non-associated character of Coulomb friction. The implications for convergence are discussed in~\S\ref{sec:stepsize}.
\end{itemize}

\subsection{Outer Iteration}
\label{sec:fbf}
 
The $A+B$ structure is the standard setting for Tseng's 
FBF splitting~\cite{tseng2000modified}. Given the current iterate $\boldsymbol{\lambda}^k$, each iteration consists of three steps.
 
\subsubsection{Evaluate the Coupling}
Compute the contact-space velocity $\mathbf{v}^k = \mathbf{W}\boldsymbol{\lambda}^k + \mathbf{v}_f$ and evaluate the 
coupling
\begin{equation}\label{eq:forward1}
  \mathbf{g}^k = B(\boldsymbol{\lambda}^k)
    = \mu\,\|\mathbf{v}^k_T\|\,\mathbf{e}_N\,.
\end{equation}
 
\subsubsection{Solve the Cone QP}
With the coupling $\mathbf{g}^k$ frozen, we solve for the regularized inclusion
\begin{equation}\label{eq:backward-inclusion}
  \mathbf{0} \in A(\bar{\boldsymbol{\lambda}}^k)
    + \mathbf{g}^k
    + \gamma^{-1}(\bar{\boldsymbol{\lambda}}^k - \boldsymbol{\lambda}^k)\,
\end{equation}
\noindent with $\gamma > 0$ the regularization parameter.

\noindent Expanding $A$ from~\eqref{eq:A} and rearranging, this becomes the optimality condition of the strongly convex cone QP:
\begin{equation}\label{eq:backward}
  \bar{\boldsymbol{\lambda}}^k
    = \operatorname*{argmin}_{\boldsymbol{\lambda}\in K_\mu}\;
      \tfrac{1}{2}\,\boldsymbol{\lambda}^\top \mathbf{W}_\gamma\,
      \boldsymbol{\lambda}
      + (\mathbf{c}^k)^\top\boldsymbol{\lambda}\,,
\end{equation}
where
\begin{equation}\label{eq:Wgamma}
  \mathbf{W}_\gamma = \mathbf{W} + \gamma^{-1}\mathbf{I}\,,
  \qquad
  \mathbf{c}^k = \mathbf{v}_f + \mathbf{g}^k
    - \gamma^{-1}\boldsymbol{\lambda}^k\,.
\end{equation}

The inclusion~\eqref{eq:backward-inclusion} is a \emph{proximal evaluation} of the operator~$A$: it finds the point $\bar{\boldsymbol{\lambda}}^k$ at which the sum of $A$ and a quadratic penalty at $\boldsymbol{\lambda}^k$ vanishes.
This is a standard construction in splitting methods~\cite{tseng2000modified}: the proximal term $\gamma^{-1}(\bar{\boldsymbol{\lambda}}^k - \boldsymbol{\lambda}^k)$ regularizes the subproblem so that it always has a unique solution, even when $\mathbf{W}$ is rank-deficient. Also, it keeps each subproblem anchored to the current iterate. Because $A$ contains only the linear response $\mathbf{W}$ and the cone constraint $N_{K_\mu}$, the proximal evaluation reduces to the cone QP~\eqref{eq:backward}: the proximal penalty becomes the $\gamma^{-1}\mathbf{I}$ term in $\mathbf{W}_\gamma$, and the frozen coupling $\mathbf{g}^k$ enters through the linear coefficient~$\mathbf{c}^k$.

This is the main computational payoff of the architecture: the inner solve is not the full nonlinear Coulomb
law~\eqref{eq:target}, but a \emph{standard cone-constrained QP} with a well-conditioned quadratic objective.
The De~Saxc\'{e} coupling has been absorbed into the linear term~$\mathbf{c}^k$ as a frozen constant, so the inner solver sees no nonlinearity beyond the cone constraint itself.

\subsubsection{Correct the Reaction}
The cone QP used the coupling $\mathbf{g}^k$ evaluated at $\boldsymbol{\lambda}^k$, but the solution
$\bar{\boldsymbol{\lambda}}^k$ has moved to a new point where $B$ may differ.
The correction compensates for this discrepancy:
\begin{equation}\label{eq:correction}
  \boldsymbol{\lambda}^{k+1}
    = \bar{\boldsymbol{\lambda}}^k
      - \gamma\bigl(B(\bar{\boldsymbol{\lambda}}^k)
      - \mathbf{g}^k\bigr)\,.
\end{equation}

If the coupling did not change ($B(\bar{\boldsymbol{\lambda}}^k) = \mathbf{g}^k$), the correction is not needed and $\boldsymbol{\lambda}^{k+1} = \bar{\boldsymbol{\lambda}}^k$.
Otherwise, it subtracts the change in $B$ scaled by~$\gamma$, keeping the iterate consistent with the exact law without requiring a second implicit solve.

\revon This correction is an explicit step, and nothing in it constrains the result to the cone. Since $B$ acts only along the normal $\mathbf{e}_N$, it moves the normal component $\lambda_N$, and a large enough correction can carry the reaction out of $K_\mu$; hence, in the limit to $\lambda_N < 0$, a pulling normal force that is not a valid Coulomb reaction. We therefore project the corrected reaction back onto the cone,
\begin{equation}\label{eq:correction-proj}
  \boldsymbol{\lambda}^{k+1}
    = \Pi_{K_\mu}\!\bigl(\bar{\boldsymbol{\lambda}}^k
      - \gamma(B(\bar{\boldsymbol{\lambda}}^k) - \mathbf{g}^k)\bigr)\,.
\end{equation}
Because the exact-law solution already lies in $K_\mu$, the projection is inactive at convergence and leaves the fixed point unchanged; it only keeps each intermediate iterate a physically admissible reaction.\revoff

The overall scheme is summarized in Algorithm~\ref{alg:fbf}.
 
\begin{algorithm}[t]
\caption{Splitting iteration for the exact reduced Coulomb law.
  The step size $\gamma$ is held fixed for simplicity;
  a safeguarded adaptive variant is described in~\S\ref{sec:stepsize}.}
\label{alg:fbf}
\begin{algorithmic}[1]
\REQUIRE $\boldsymbol{\lambda}^0 \in K_\mu$,\; step size $\gamma > 0$,\;
  tolerance $\varepsilon$
\FOR{$k = 0, 1, 2, \ldots$}
  \STATE $\mathbf{v}^k \leftarrow \mathbf{W}\boldsymbol{\lambda}^k + \mathbf{v}_f$
    \hfill\COMMENT{contact velocity}
  \STATE $\mathbf{g}^k \leftarrow
    \mu\,\|\mathbf{v}^k_T\|\,\mathbf{e}_N$
    \hfill\COMMENT{De Saxc\'{e} coupling}
  \STATE $\bar{\boldsymbol{\lambda}}^k \leftarrow
    \text{solve cone QP~\eqref{eq:backward}}$
    \hfill\COMMENT{cone QP}
  \STATE $\bar{\mathbf{v}}^k \leftarrow
    \mathbf{W}\bar{\boldsymbol{\lambda}}^k + \mathbf{v}_f$
  \STATE $\bar{\mathbf{g}}^k \leftarrow
    \mu\,\|\bar{\mathbf{v}}^k_T\|\,\mathbf{e}_N$
    \hfill\COMMENT{coupling at new point}
  \STATE \revon$\boldsymbol{\lambda}^{k+1} \leftarrow
    \Pi_{K_\mu}\!\bigl(\bar{\boldsymbol{\lambda}}^k
    - \gamma(\bar{\mathbf{g}}^k - \mathbf{g}^k)\bigr)$\revoff
    \hfill\COMMENT{correction}
  \STATE \textbf{if} $r_c(\boldsymbol{\lambda}^{k+1}) < \varepsilon$
    \textbf{then stop}
    \hfill\COMMENT{Coulomb residual~\eqref{eq:rc}}
\ENDFOR
\end{algorithmic}
\end{algorithm}

\subsection{The Inner Cone Subproblem}
\label{sec:inner-solve}
The cone QP~\eqref{eq:backward} is a strongly convex quadratic program over a product of per-contact Coulomb cones.
Since $\mathbf{W}_\gamma = \mathbf{W} + \gamma^{-1}\mathbf{I}$ is positive definite, the problem is well-posed regardless of the rank of $\mathbf{W}$. This is a direct benefit of the proximal regularization introduced by the outer iteration: even for underdetermined contact configurations where $\mathbf{W}$ is singular, the inner subproblem always has a unique solution.

The outer iteration is agnostic to how the cone QP is solved: it requires only the minimizer $\bar{\boldsymbol{\lambda}}^k$. Any solver that handles strongly convex QPs over second-order cones can be used, including block Gauss--Seidel sweeps~\cite{ascher2011first}, accelerated projected gradient methods, first-order cone solvers such as SCS~\cite{SCS}, or interior-point solvers such as Clarabel~\cite{Clarabel}. This modularity is one of the architectural features of the splitting: the inner solver can be chosen or replaced based on the contact count, sparsity pattern, or available hardware, without changing the outer loop.

Across outer iterations, consecutive subproblems differ only in the linear coefficient $\mathbf{c}^k$, so the previous solution $\bar{\boldsymbol{\lambda}}^{k-1}$ provides an informed starting point for $\bar{\boldsymbol{\lambda}}^k$. Any inner solver that accepts an initial guess benefits from this warm start. In our implementation we use a matrix-free block Gauss--Seidel sweep as the default inner solver; details are given in~\S\ref{sec:implementation}.

\subsection{Step-Size Control}
\label{sec:stepsize}
The parameter $\gamma$ in Algorithm~\ref{alg:fbf} is an \emph{algorithmic} step size of the outer splitting iteration; it is unrelated to the physical time step $h$. 
It has a dual role: $\gamma^{-1}$ regularizes the cone-constrained subproblem solved in the middle stage, and $\gamma$ steps the explicit correction of the reaction in the final stage. The same value must appear in both places for the splitting structure of \S\ref{sec:algorithm} to be preserved.

\paragraph*{A base step size from Lipschitz analysis.}
The explicit coupling $B$ defined in~\eqref{eq:B} is Lipschitz continuous in $\boldsymbol{\lambda}$: its variation is controlled by the friction coefficient and by the global scale of the Delassus operator $\mathbf{W}$. Since
  $\mathbf{W}$ is symmetric positive semidefinite in the rigid-body setting, $\|\mathbf{W}\boldsymbol{\lambda}\| \leq \lambda_{\max}(\mathbf{W})\,\|\boldsymbol{\lambda}\|$, and a direct
  computation yields
\begin{equation}
\| B(\boldsymbol{\lambda}_1) - B(\boldsymbol{\lambda}_2) \|
\;\leq\;
\mu_{\max} \, \lambda_{\max}(\mathbf{W}) \, \| \boldsymbol{\lambda}_1 - \boldsymbol{\lambda}_2 \|,
\label{eq:Lipschitz}
\end{equation}
where $\mu_{\max} = \max_i \mu_i$. This bound motivates a conservative base step size
\begin{equation}
\gamma_{\mathrm{safe}}
\;=\;
\frac{c_\gamma}{\mu_{\max} \, \hat\lambda_{\max}(\mathbf{W})},
\qquad c_\gamma = 0.5,
\label{eq:gamma_safe}
\end{equation}
where $\hat\lambda_{\max}(\mathbf{W})$ is an estimate of the largest eigenvalue of $\mathbf{W}$ obtained from a short power iteration (ten matrix--vector products in our implementation), and $c_\gamma < 1$ is a safety factor.

\paragraph*{Why the choice of $\gamma$ matters.}
A small $\gamma$ yields conservative outer updates and strong proximal regularization of the cone subproblem: the iteration is stable but slow to build up frictional support. A large $\gamma$ weakens the regularization and makes the explicit correction more aggressive. This accelerates convergence when all contacts remain well inside either the sticking or the sliding regime, but can destabilize the outer iteration near the sticking--sliding boundary. Therefore, this step size is not a free parameter to be easily tuned by hand. It determines success or failure on stacking and arch benchmarks, which motivates the safeguarded rule described below.

\paragraph*{Non-monotonicity and Tseng's mechanism.}
The three-stage iteration of Algorithm~\ref{alg:fbf} has the structure of Tseng's
method~\cite{tseng2000modified}. 
However, the main theorem in that work cannot be directly applied here because the coupling $B(\lambda)$
is not monotone in general.
This is a direct consequence of the non-associated character of Coulomb friction, not an artifact of our formulation, and it is shared by every exact reduced-law method we are aware of. 
Therefore, we adopt the
\emph{adaptive step-size mechanism}~\cite[eq.~(2.4)]{tseng2000modified} as a practical safeguard, and establish convergence empirically.

\paragraph*{Safeguarded adaptive rule.}
At outer iteration $k$, let $\gamma_k$ be a trial step size and let $\bar{\boldsymbol{\lambda}}^k$ denote the intermediate iterate returned by the cone subproblem~\eqref{eq:backward}. Define the local variation ratio
\begin{equation}
\rho_k
\;=\;
\gamma_k \,
\frac{\| B(\bar{\boldsymbol{\lambda}}^k) - B(\boldsymbol{\lambda}^k) \|}{\| \bar{\boldsymbol{\lambda}}^k - \boldsymbol{\lambda}^k \|}.
\label{eq:rho}
\end{equation}
We accept the trial step size if $\rho_k \leq \theta$ for a fixed threshold $\theta \in (0,1)$; otherwise, we 
shrink $\gamma_k \leftarrow \beta \gamma_k$ with $\beta \in (0,1)$, and re-solve the cone subproblem. The correction $\boldsymbol{\lambda}^{k+1} = \bar{\boldsymbol{\lambda}}^k - \gamma_k (B(\bar{\boldsymbol{\lambda}}^k) - B(\boldsymbol{\lambda}^k))$ is applied only with an accepted $\gamma_k$. The condition $\rho_k \leq \theta$ 
bounds the local variation of the explicit coupling~$B$ by a fraction $\theta$ of the variation of $\boldsymbol{\lambda}$ across the proximal step. We use $\theta = 0.9$ and $\beta = 0.7$ throughout.

\paragraph*{Asymmetric growth policy.}
The rule above only shrinks $\gamma$. 
To increase it when conditions allow such growth, we use an asymmetric policy. Within a single outer solve, once the step size has been shrunk it is not allowed to grow again. Across outer solves, $\gamma$ is allowed to increase
but is capped at $\gamma_{\mathrm{safe}}$~\eqref{eq:gamma_safe}. In practice $\gamma$ is a conservative starting estimate that is reduced on demand when the local variation ratio $\rho_k$ exceeds the acceptance threshold, and slowly restored when consecutive simulation steps produce no rejections. This policy is not required by any convergence theorem; it reflects the empirical observation that unconditional growth destabilizes the iteration in regimes where the sticking/sliding boundary is active, while the conservative base step $\gamma_{\mathrm{safe}}$ is reliable elsewhere.

\subsection{Stopping Criterion}
\label{sec:stopping}
We need a scalar measure that vanishes if and only if the current iterate $\boldsymbol{\lambda}$ satisfies the exact cone complementarity~\eqref{eq:soccp}.
Recall that the target law requires
\[
  K^*_\mu \ni \tilde{\mathbf{v}} \;\perp\; \boldsymbol{\lambda} \in K_\mu\,,
\]
i.e.\ the augmented velocity lies in the dual cone, the reaction lies in the primal cone, and the two are orthogonal.
For any closed convex cone~$K$, this complementarity is equivalent to the fixed-point condition
\begin{equation}\label{eq:fixedpoint}
  \boldsymbol{\lambda}
    = \Pi_{K_\mu}\!\bigl(\boldsymbol{\lambda}
      - \tilde{\mathbf{v}}(\boldsymbol{\lambda})\bigr)\,,
\end{equation}
where $\Pi_{K_\mu}$ denotes projection onto~$K_\mu$.
To see why, note that $-\tilde{\mathbf{v}} \in N_{K_\mu}(\boldsymbol{\lambda})$ (the normal cone) is just a restatement of the complementarity, and the projection characterization of the normal cone gives $\boldsymbol{\lambda} = \Pi_{K_\mu}(\boldsymbol{\lambda} - \tilde{\mathbf{v}})$ if and only if $-\tilde{\mathbf{v}} \in N_{K_\mu}(\boldsymbol{\lambda})$.

\revon The gap in this fixed-point equation defines the \emph{natural-map residual}
\begin{equation}\label{eq:residual}
  \mathcal{R}(\boldsymbol{\lambda})
    = \bigl\|\boldsymbol{\lambda}
    - \Pi_{K_\mu}\!\bigl(\boldsymbol{\lambda}
      - \tilde{\mathbf{v}}(\boldsymbol{\lambda})\bigr)\bigr\|\,,
\end{equation}
which has the correct zero set: $\mathcal{R}(\boldsymbol{\lambda}) = 0$ if and only if the exact Coulomb law is satisfied, because the projection identity $\boldsymbol{\lambda} = \Pi_{K_\mu}(\boldsymbol{\lambda} - \alpha\,\tilde{\mathbf{v}})$ holds for any $\alpha > 0$.
As a numerical stopping threshold, however, the choice $\alpha = 1$ in~\eqref{eq:residual} is poorly scaled: it subtracts the augmented velocity $\tilde{\mathbf{v}}$ (units of velocity) from the reaction $\boldsymbol{\lambda}$ (units of impulse) inside a single projection, so whichever side carries the larger magnitude dominates. On contact-rich scenes, this inflates the residual by the impulse-to-velocity scale ratio rather than by the true complementarity violation.

We therefore declare convergence on a dimensionless residual that scores the three conditions of~\eqref{eq:soccp} separately. Fixing an impulse scale $s_r$ and a velocity scale $s_u$,
\begin{equation}\label{eq:scales}
  s_r = \max\!\bigl(\|\boldsymbol{\lambda}^0\|,\; \|\mathbf{v}_f\|/\hat\lambda_{\max}(\mathbf{W}),\; \epsilon_0\bigr),
  \quad
  s_u = \max\!\bigl(\|\tilde{\mathbf{v}}^0\|,\; \|\mathbf{v}_f\|,\; \epsilon_0\bigr),
\end{equation}
held constant over the outer loop (with floor $\epsilon_0 = 10^{-12}$, and $\hat\lambda_{\max}(\mathbf{W})$ reusing the power-iteration estimate of~\eqref{eq:gamma_safe}), we define
\begin{equation}\label{eq:split-residual}
\begin{aligned}
  \varepsilon_{\mathrm{force}} &= \|\boldsymbol{\lambda} - \Pi_{K_\mu}(\boldsymbol{\lambda})\| \,/\, s_r,\\
  \varepsilon_{\mathrm{vel}}   &= \|\tilde{\mathbf{v}} - \Pi_{K^*_\mu}(\tilde{\mathbf{v}})\| \,/\, s_u,\\
  \varepsilon_{\mathrm{gap}}   &= |\tilde{\mathbf{v}}^\top\boldsymbol{\lambda}| \,/\, (s_r\,s_u),
\end{aligned}
\end{equation}
measuring primal feasibility ($\boldsymbol{\lambda}\in K_\mu$), dual feasibility ($\tilde{\mathbf{v}}\in K^*_\mu$), and orthogonality, respectively, and stop when
\begin{equation}\label{eq:rc}
  r_c(\boldsymbol{\lambda})
    = \max\bigl(\varepsilon_{\mathrm{force}},\, \varepsilon_{\mathrm{vel}},\, \varepsilon_{\mathrm{gap}}\bigr)
    < \varepsilon\,.
\end{equation}
Each component is unit-consistent, so $r_c$ is dimensionless and the threshold $\varepsilon$ carries the same meaning across scenes, from a single-contact ball to the masonry arch.
The dual cone $K^*_\mu$ is the Coulomb cone with reciprocal coefficient $1/\mu$.
Because the inner cone solve always returns a primal-feasible $\boldsymbol{\lambda}\in K_\mu$, the term $\varepsilon_{\mathrm{force}}$ vanishes in practice and $r_c$ is governed by dual feasibility and the complementarity gap.
Evaluating $r_c$ requires one matrix--vector product $\mathbf{W}\boldsymbol{\lambda}$ (to form $\tilde{\mathbf{v}}$) and one cone projection per contact, the same cost as the natural map.
We use $\varepsilon = 10^{-6}$: since the pipeline is single precision, $\varepsilon_{\mathrm{vel}}$ and $\varepsilon_{\mathrm{gap}}$ carry a rounding floor of order $\epsilon_{32}\sqrt{3 n_c} \approx 2\times 10^{-6}$ at $n_c \approx 108$ (single-precision unit roundoff $\epsilon_{32}\approx 1.2\times10^{-7}$), so a tighter threshold would test rounding noise rather than convergence.\revoff

\subsection{Implementation}
\label{sec:implementation}
 
We implement the solver on top of NVIDIA Warp~\cite{warp2022} and the Newton physics engine~\cite{newton2025}. Newton is used to handle scene setup, collision detection, and broad/narrow-phase contact generation. 
Our solver replaces Newton's built-in contact resolution with the FBF pipeline operating entirely in reduced contact space. The implementation has the following structure.
 
\subsubsection{Contact Frontend}
The collision pipeline gives contact points, normals, and penetration depths in body-local coordinates.
We transform these to world space, build deterministic contact frames, compute per-contact friction coefficients, and apply Baumgarte stabilization to the normal component of the free velocity.
 
\subsubsection{Matrix-Free Operators}
The Delassus operator $\mathbf{W} = \mathbf{J}\mathbf{M}^{-1}\mathbf{J}^\top$ is never formed or factorized as a dense matrix. Both the outer FBF loop and the inner solver access $\mathbf{W}$ only through matrix vector products and per-contact diagonal blocks:

\begin{itemize}
\item
Each product $\mathbf{W}\boldsymbol{\lambda}$ is evaluated
as a three-stage pipeline--scatter reactions to body wrenches ($\mathbf{J}^\top$), apply the diagonal inverse mass ($\mathbf{M}^{-1}$), and gather contact velocities ($\mathbf{J}$). These are all parallelized over contacts and bodies.

\item
The inner block Gauss--Seidel solver requires only the $3{\times}3$ diagonal blocks $\mathbf{W}_{ii} = \mathbf{J}_i\mathbf{M}^{-1}\mathbf{J}_i^\top$, which are computed once per contact from local Jacobian and mass data, and the off-diagonal coupling $\mathbf{W}_{ij}\boldsymbol{\lambda}_j$ between contacts sharing a body, which is evaluated on the fly during each sweep.
\end{itemize}

The only matrix inversions are the per-body $\mathbf{M}^{-1}$ and the per-contact $3{\times}3$ blocks in the Gauss--Seidel solve. The cost therefore scales linearly with the number of contacts and the contact-body adjacency, not with a dense factorization of the full $3n_c \times 3n_c$ system.
 
\subsubsection{Warm Starting}
When the contact configuration is unchanged between time steps, we reuse the previous solution as the initial $\boldsymbol{\lambda}^0$, which typically halves the iteration count.
 
\subsubsection{Parallelism}
The coupling evaluation and correction steps are embarrassingly parallel over contacts. The inner cone QP uses a sequential Gauss--Seidel sweep, but each per-contact 3D solve is independent given the current values of its neighbors. For large contact counts, the inner solver can be replaced by a GPU-native cone solver without changing the outer loop.

\section{Results}
\label{sec:results}














We evaluate FBF on a sequence of rigid-body benchmarks that probe stick--slip transitions, normal--tangential coupling under sliding, and static frictional support in multi-contact structures. On every scene we compare against two representative solvers that are available within the same Newton framework~\cite{newton2025}: Kamino~\cite{kamino2026, tsounis2025solving}, a very recent, industrial strength physics solver that targets exact Coulomb friction, and MuJoCo~\cite{todorov2012mujoco}, a widely used engine based on the convex-relaxed CCP formulation. Kamino therefore serves as a reference exact-law solver, while MuJoCo exposes the artifacts introduced by the CCP formulation. Where a closed-form solution is available, we include it as a third reference.

\subsection{Cube on Incline}
\label{sec:incline}

\begin{figure}[t]
  \centering
  \includegraphics[width=\linewidth]{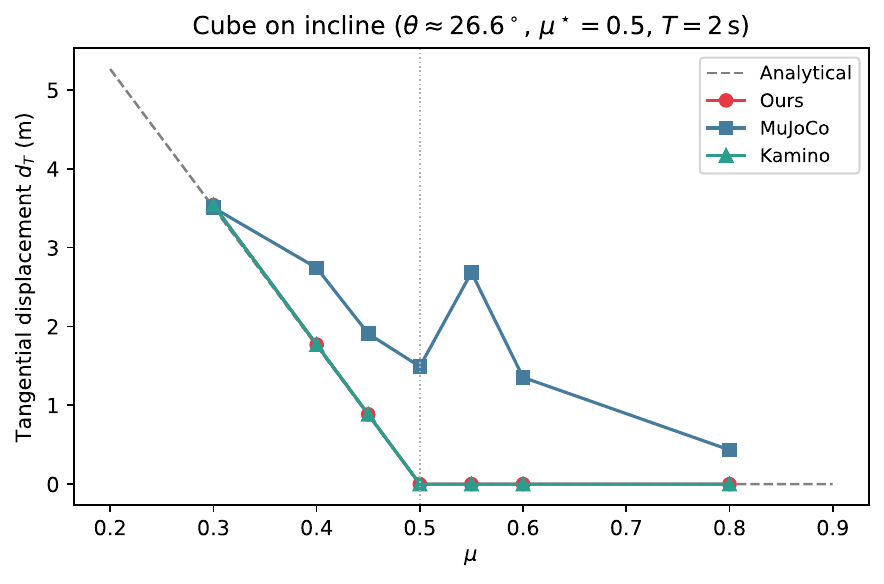}
  \caption{Tangential displacement $d_T(\mu)$ after $T=2\,\mathrm{s}$ for a cube on an inclined plane at $\theta\approx 26.6^\circ$. FBF (red) and Kamino (green) follow the analytical solution (dashed) across the full sweep, with the correct transition from sliding to sticking at $\mu^\star=0.5$. MuJoCo (blue) overshoots in the sliding regime and exhibits spurious non-monotone drift above the threshold.}
  \label{fig:incline-sweep}
\end{figure}

\begin{figure}[t]
  \centering
  \includegraphics[width=\linewidth]{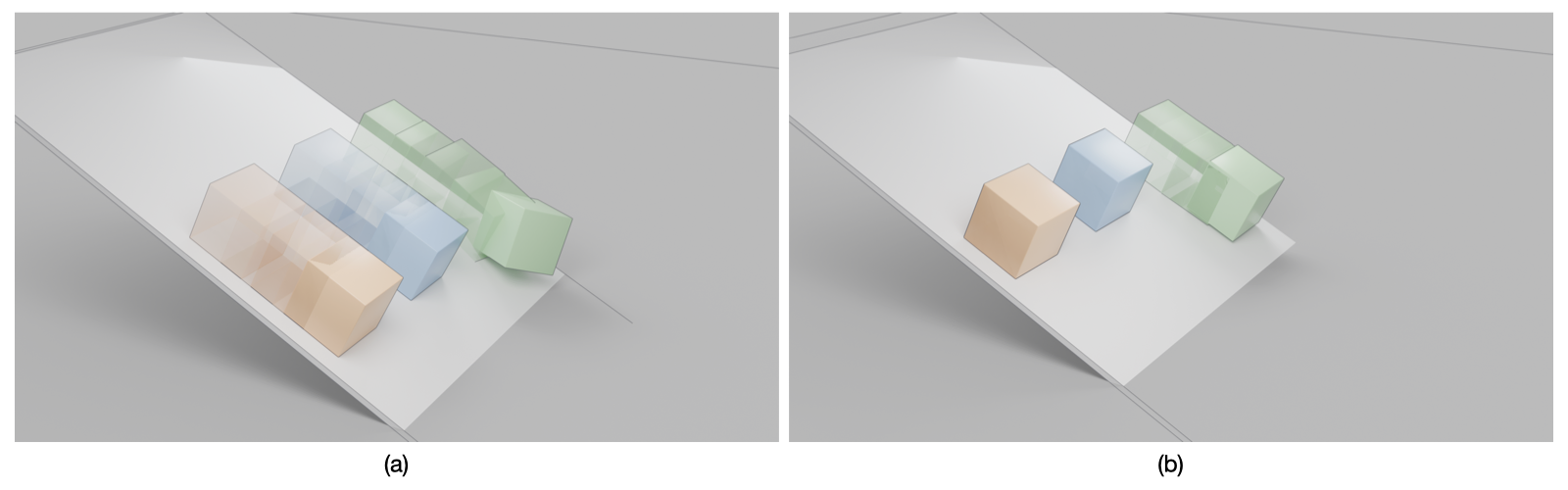}
  \caption{Snapshots of the cube on an inclined plane, showing FBF (orange), Kamino (blue), and MuJoCo (green). (a) $\mu=0.4$, below threshold: FBF and Kamino slide together down the plane in sustained contact; MuJoCo drifts farther. (b) $\mu=0.5$, at threshold: FBF and Kamino remain stuck; MuJoCo continues to creep downhill.}
  \label{fig:incline-snapshots}
\end{figure}

We fix the slope at $\theta=\arctan(0.5)\approx 26.6^\circ$ so that
$\mu^\star=0.5$, sweep the friction coefficient across this threshold, and
simulate each case for $T=2\,\mathrm{s}$ with $\Delta t = 1/60\,\mathrm{s}$.
Figure~\ref{fig:incline-sweep} reports the tangential displacement $d_T(\mu)$.
FBF and Kamino track the analytical reference on both sides of the threshold:
constant-acceleration sliding below $\mu^\star$ and exact stick at and above
it. MuJoCo fails on both sides. In the sliding regime it overshoots the
analytical curve, and above the threshold it exhibits nonzero, non-monotone
drift, with the largest deviation near $\mu=0.55$. Figure~\ref{fig:incline-snapshots}
shows the corresponding 3D configurations at $\mu=0.4$ and $\mu=0.5$: FBF and
Kamino remain in stable contact with the plane across the sweep, while
MuJoCo's cube loses sustained contact and drifts, reflecting the normal lifting artifact in the convex-relaxed CCP formulation.

\subsection{Ball with Backspin}
\label{sec:backspin}
A uniform solid sphere of radius \(r = 0.25\,\mathrm{m}\) is launched on a
horizontal plane with \(v_0 = 4\,\mathrm{m/s}\), \(\omega_0 = -200\,\mathrm{rad/s}\),
and \(\mu = 0.5\), at \(\Delta t = 1/60\,\mathrm{s}\). Under exact Coulomb
friction, the backspin reverses the ball's horizontal motion and drives it
to a pure-rolling rest with limiting velocity
\begin{equation}
\label{eq:backspin-analytical}
v_\infty = \frac{5v_0 + 2r\omega_0}{7},
\qquad
\omega_\infty = \frac{v_\infty}{r},
\end{equation}
giving \(v_\infty = -11.429\,\mathrm{m/s}\) and
\(\omega_\infty = -45.71\,\mathrm{rad/s}\) for these parameters.

\begin{figure}[t]
  \centering
  \includegraphics[width=\linewidth]{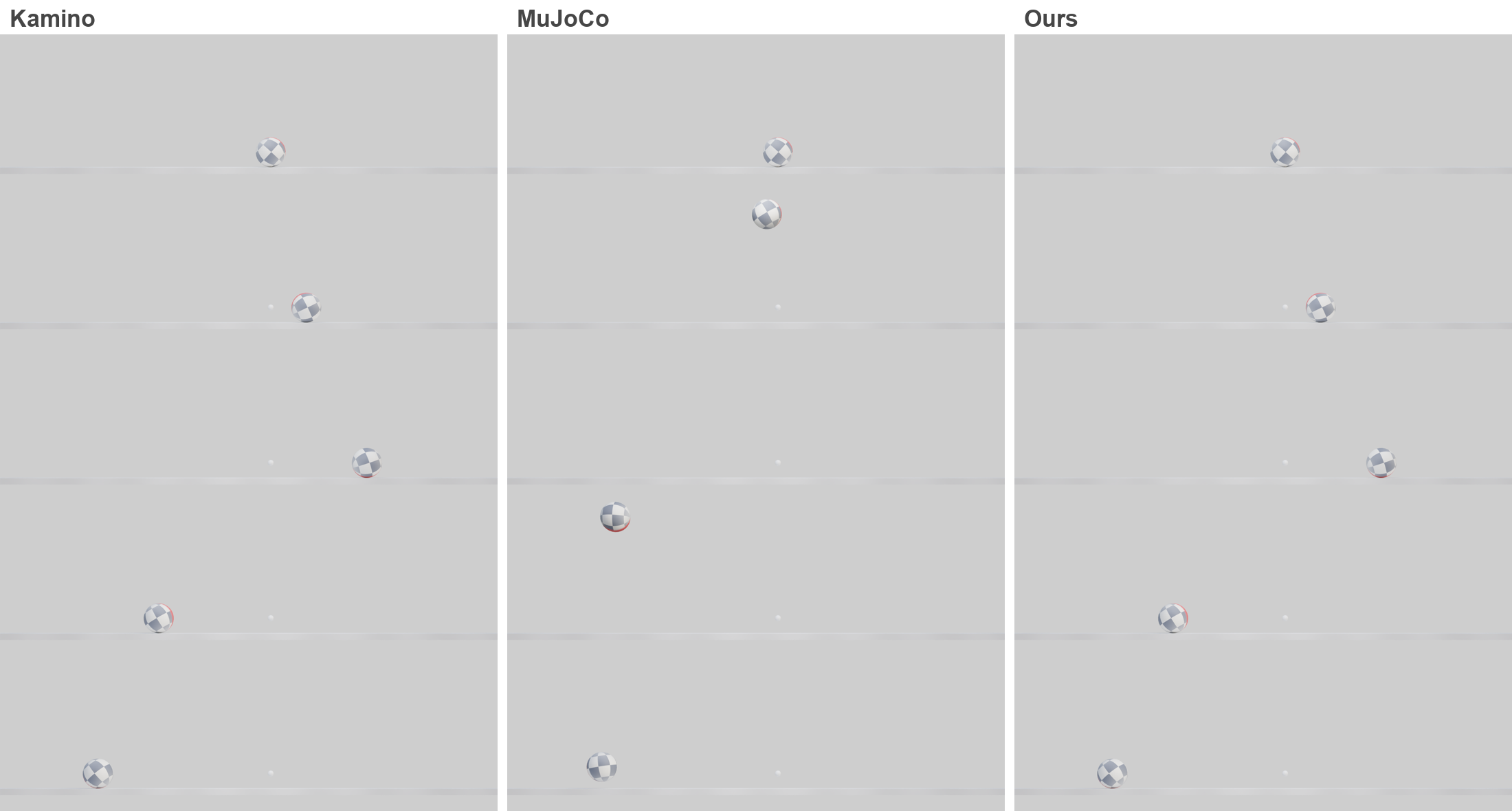}
  \caption{Backspin-ball benchmark. Each column shows one solver at the same five instants $t=0$, $0.17$, $0.83$, $2.0$, and $2.17\,\mathrm{s}$. Kamino (left) and FBF (right) track the same trajectory: the ball lands, advances briefly, then reverses under the backspin and rolls back to rest. MuJoCo (middle) instead loses contact and is ejected vertically off the plane, rising out of frame (the empty third row) before falling back to the ground.}
  \label{fig:backspin}
\end{figure}

Figure~\ref{fig:backspin} shows the ball's trajectory for FBF, Kamino, and
MuJoCo. FBF and Kamino agree closely: the ball descends onto the plane,
travels briefly forward, then reverses under the backspin and comes to rest
behind the launch point, as the exact law predicts. MuJoCo produces a
qualitatively different result: the ball is ejected vertically off the plane
within a few time steps and thrown clear of the surface in a long ballistic
arc before falling back. This is the backspin signature of the missing De~Saxc\'{e} coupling:
large tangential velocity at contact drives a strong normal response in the
exact law, and a convex-relaxed CCP solver cannot reproduce it without
losing sustained contact.

\subsection{Turntable}
\label{sec:turntable}
\begin{figure}[t]
  \centering
  \includegraphics[width=\linewidth]{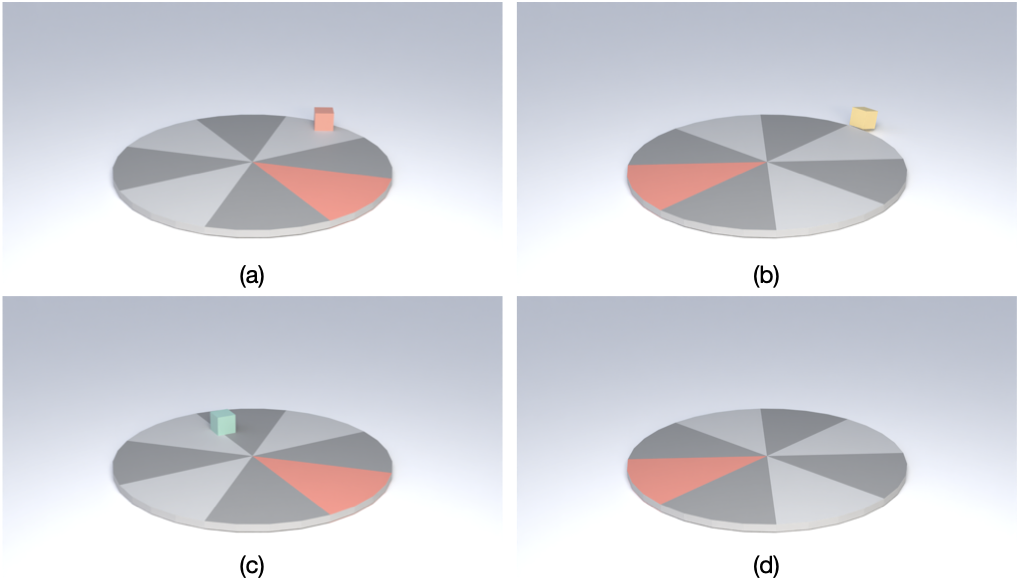}
  \caption{Cube on a rotating turntable across friction coefficients and spin rates. (a) $\mu=0.2$, $\omega=2$: ejected. (b) $\mu=0.2$, $\omega=5$: ejected. (c) $\mu=0.5$, $\omega=2$: captured and co-rotating with the disk. (d) $\mu=0.5$, $\omega=5$: ejected.}
  \label{fig:turntable}
\end{figure}

We place a cube on a rotating turntable whose angular speed is ramped smoothly
from rest, as a check of capture and ejection under centripetal
loading. Figure~\ref{fig:turntable} shows four runs across friction
coefficients and spin rates. At low friction (\(\mu=0.2\)), the cube is
thrown off at both tested angular speeds. At higher friction (\(\mu=0.5\)),
the cube is captured and approximately co-rotates at the lower spin rate
(\(\omega=2\)) but slips outward and is ejected at the higher rate
(\(\omega=5\)). The sweep illustrates the expected trend that larger friction
enlarges the range of captured speeds, while higher spin rates drive outward
slip and eventual loss of contact.

\subsection{Painlev\'{e} Box}
\label{sec:painleve}
\begin{figure}[t]
    \centering
    \includegraphics[width=\linewidth]{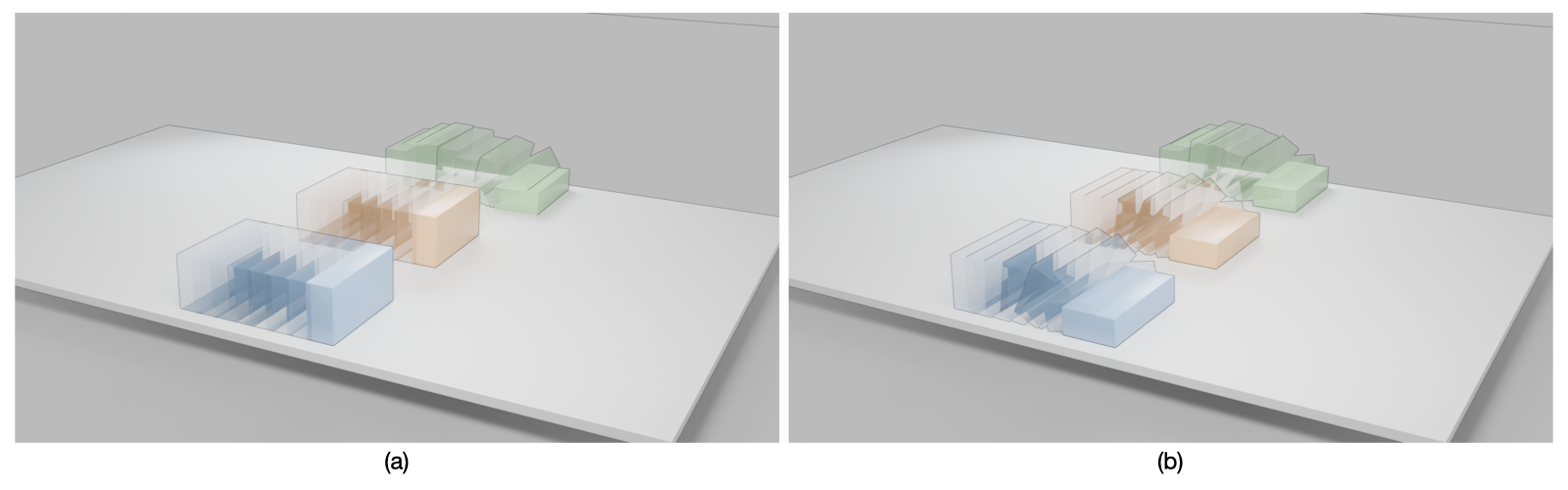}
    \caption{Painlev\'{e} box trajectories showing FBF (orange), Kamino (blue), and MuJoCo (green). (a) $\mu=0.5$: FBF and Kamino slide forward without tipping, while MuJoCo travels farther and tips forward. (b) $\mu=0.55$: FBF and Kamino travel a shorter distance before tipping, while MuJoCo again travels farther and tips forward. FBF and Kamino agree closely in both cases.}
    \label{fig:painleve}
\end{figure}

We place a box on a rough horizontal plane and sweep the friction
coefficient near the quasi-static tipping threshold, using the example as a
visual check of the slide--to--tumble transition. Figure~\ref{fig:painleve}
shows representative runs at $\mu=0.5$ and $\mu=0.55$. At $\mu=0.5$, both
FBF and Kamino slide and then come to rest without tumbling, while MuJoCo
shows an initial jump and subsequently tumbles. When the friction is
increased to $\mu=0.55$, both FBF and Kamino tumble, indicating that the
system has crossed the transition, whereas MuJoCo again exhibits an initial
jump before tumbling. Although no closed-form solution is available for the
full transient motion, the comparison shows that FBF and Kamino produce
consistent qualitative behavior across the threshold, while MuJoCo displays
a premature separation event near the transition.

\subsection{House of Cards}
\label{sec:cardhouse}

We test FBF on a rigid house-of-cards benchmark, a friction-dependent
structure whose stability relies on accurate local coupling between normal
support and tangential friction. Small errors in this balance accumulate into
outward creep or premature collapse, making the example a sensitive test of
static friction resolution.

\begin{figure}[t]
  \centering
  \includegraphics[width=\linewidth]{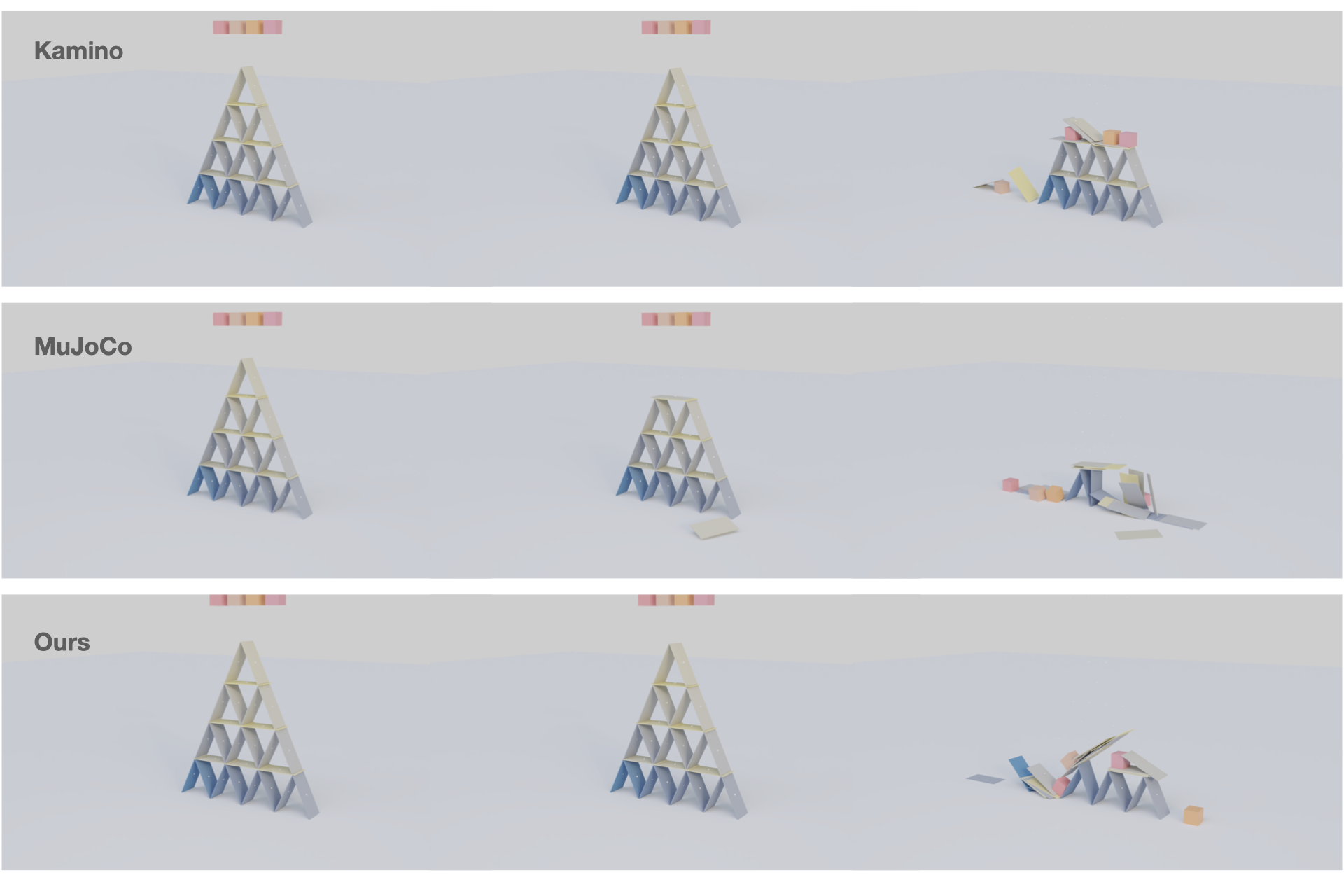}
  \caption{House of cards: $26$ plates across four levels, $\mu=0.8$. Each row shows one solver at three moments: initial settled state (left), continued standing at $t=6.7\,\mathrm{s}$ (middle), and post-impact state at $t=10\,\mathrm{s}$ after four projectiles strike one side (right). FBF and Kamino remain standing through the rest phase and fail locally near the impact. MuJoCo loses the standing configuration during the rest phase, before any projectile arrives.}
  \label{fig:cardhouse}
\end{figure}

The scene consists of $26$ thin rigid plates assembled into a four-level
pyramidal house of cards with $\mu = 0.8$ and $\Delta t = 1/60\,\mathrm{s}$.
After gravity settling, we let the structure stand and then launch four
projectiles into one side. Figure~\ref{fig:cardhouse} compares FBF, Kamino,
and MuJoCo across three moments: initial rest, continued standing after
$6.7\,\mathrm{s}$, and the post-impact state at $10\,\mathrm{s}$. FBF and
Kamino keep the stack standing through the rest phase with no visible creep,
and after the projectile impact they produce a localized failure that
propagates through neighboring cards as contacts are reconfigured. MuJoCo
fails to maintain the standing configuration: the structure drifts apart
during the rest phase before any projectile arrives. This benchmark
therefore tests not only whether the structure can stand, but whether the
solver preserves the local frictional support mechanism on which the
structure depends.

\subsection{Masonry Arch}
\label{sec:arch}

We further evaluate FBF on a masonry arch assembled from rigid voussoirs.
Unlike the house of cards, which tests local support, arch stability depends
on global force transmission across many frictional contacts, and small local
errors in any of them can destroy an otherwise valid equilibrium.

\begin{figure}[t]
  \centering
  \includegraphics[width=\linewidth]{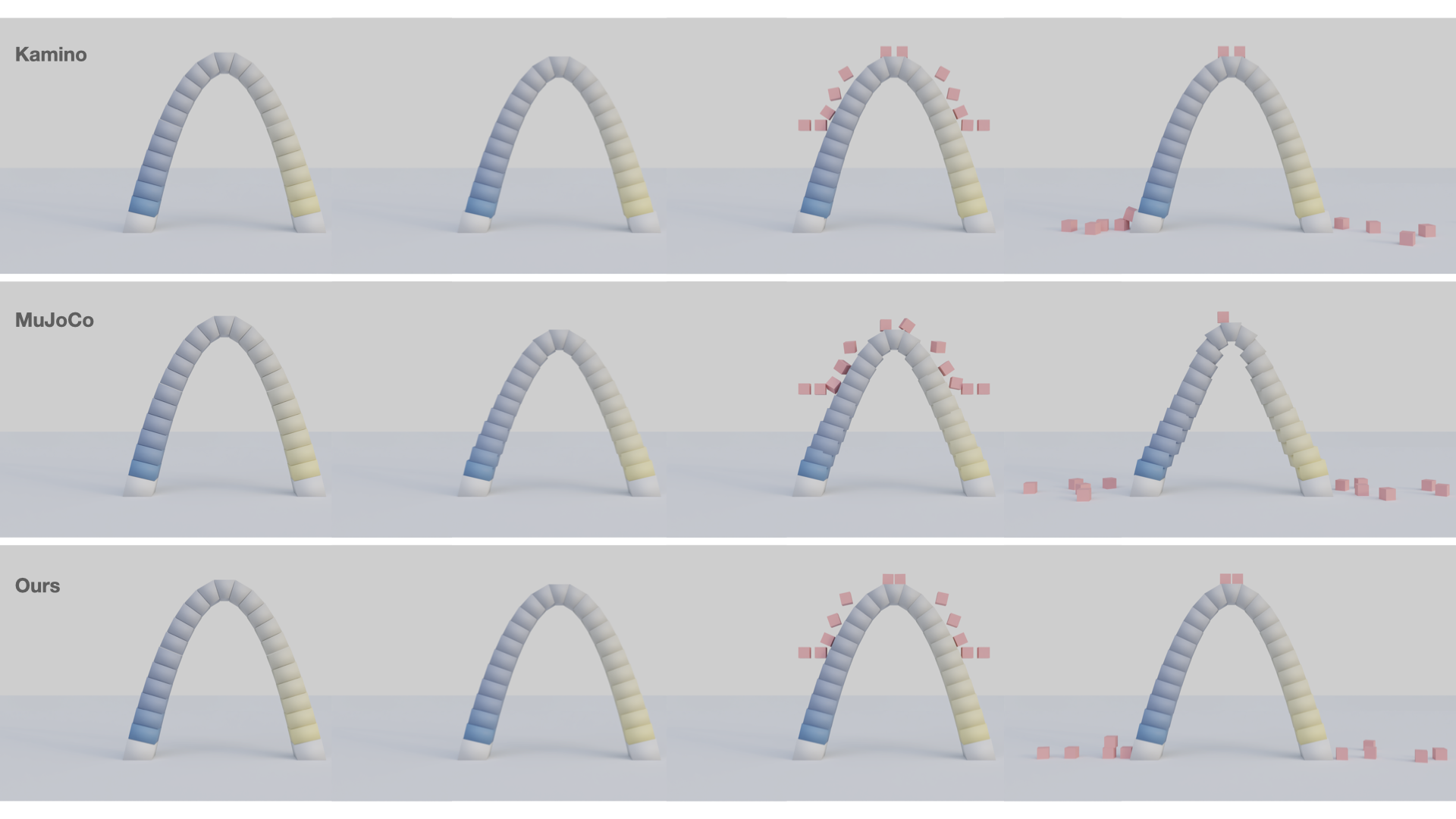}
  \caption{Masonry arch under projectile impact at the crown, shown for Kamino (top), MuJoCo (middle), and FBF (bottom) across four moments: gravity settling, continued rest, impact, and the post-impact state. FBF and Kamino settle to the same equilibrium, withstand the impact, and preserve the arch geometry. MuJoCo slumps before impact and collapses at the crown afterward.}
  \label{fig:arch}
\end{figure}

The scene is a semicircular arch of \revon $25$ voussoirs, with the two end stones pinned as abutments at the springing; all $23$ interior stones are dynamic and held in place purely by contact and friction.\revoff{} After settling under gravity, we launch a cluster of small projectiles at the crown and let the arch respond.
Figure~\ref{fig:arch} compares FBF, Kamino, and MuJoCo across the same four
moments: gravity-settled rest, continued standing, projectile impact at the
crown, and the post-impact state. FBF and Kamino settle to the same
equilibrium and remain standing through the impact, absorbing the projectile
through local rearrangement near the crown while the global arch geometry is
preserved. MuJoCo already loses the settled configuration before impact,
with visible slumping of the piers, and the arch collapses at the crown once
the projectile arrives. Together with the house of cards, this example
probes the two regimes in which static friction error is most visible: local
support in the former, and global frictional transmission in the latter.

\begin{figure}[t]
  \centering
  \includegraphics[width=\linewidth]{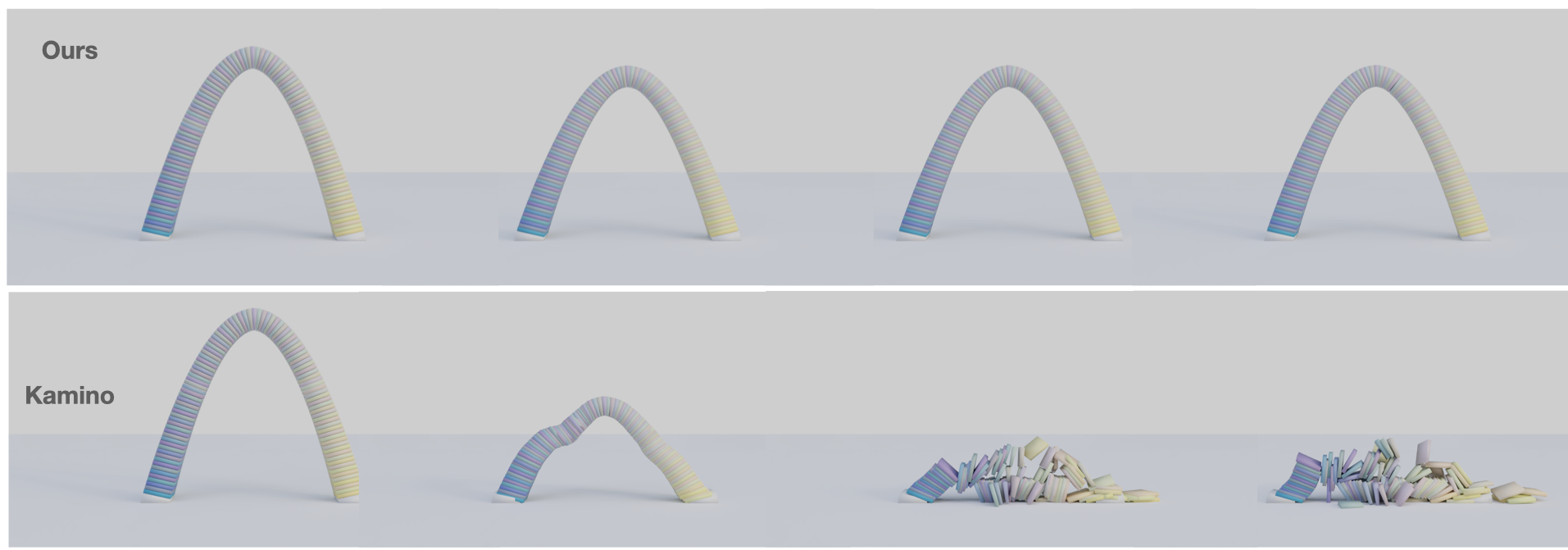}
  \caption{Masonry arch with 101 stones: our solver keeps the arch balanced, whereas Kamino fails under the same setup.}
  \label{fig:scale}
\end{figure}

We also test a 101-stone arch against Kamino, where our solver keeps the structure standing over a long run while Kamino fails (Figure~\ref{fig:scale}).

\revon
\subsection{Computational Cost}
\label{sec:performance}

We measure wall-clock cost under a residual tolerance matched across the three solvers, stopping FBF, MuJoCo, and Kamino at $\varepsilon=10^{-6}$, with FBF capped at $200$ outer iterations under the adaptive step size of~\S\ref{sec:stepsize} and a fixed budget of $10$ inner block Gauss--Seidel sweeps per outer step, raised to $30$ on the arch. On the two contact-rich baselines, where the contact graph carries force across the whole structure, the exact-law solver is competitive with the industrial-strength Kamino and an order of magnitude faster than the convex-relaxed MuJoCo (Table~\ref{tab:perf}). FBF steps the house of cards in $199\,\mathrm{ms}$ against MuJoCo's $1.7\,\mathrm{s}$, while Kamino's default blocked factorization over-sizes the chained card--card contact graph, so on this scene it runs only through its slower matrix-free path (Appendix~\ref{app:performance}). On the masonry arch FBF is more than twenty times faster than MuJoCo and within a factor of three of Kamino, the baseline on which Kamino's default factorization runs to completion.

\begin{table}[t]
  \centering
  \caption{Per-step wall time on the two contact-rich baselines at matched residual tolerance $\varepsilon=10^{-6}$ ($\mu=0.8$). FBF time is the mean wall time per step under a $200$-iteration outer budget with $10$ inner block Gauss--Seidel sweeps per outer step ($30$ on the arch); \texttt{outer} is the median outer-iteration count. The MuJoCo and Kamino columns are relative to FBF. Kamino's default blocked factor exceeds the array-size limit on the house of cards, where it runs only through its slower matrix-free path; that factor is a lower bound measured before the run was stopped.}
  \label{tab:perf}
  \small
  \begin{tabular}{l c c r c c}
    \hline
    benchmark & $n_c$ & outer & FBF (ms) & MuJoCo & Kamino \\
    \hline
    house of cards & $214$ & $5$   & $199$ & $8.7\times$  & $\ge\!50\times$ \\
    masonry arch   & $100$ & $200$ & $595$ & $23.3\times$ & $0.38\times$ \\
    \hline
  \end{tabular}
\end{table}

The advantage widens with scale. On the 101-stone arch of Figure~\ref{fig:scale} and a ten-level house of cards, FBF is the only solver that completes both runs faithfully, as MuJoCo finishes neither scene within many times FBF's wall time and Kamino, though its matrix-free path now runs the ten-level house of cards, still integrates a contact-free trajectory on the arch. On the small single- and few-contact scenes the three solvers sit within a few milliseconds of one another, with MuJoCo modestly ahead where there is sustained contact and FBF ahead on the contact-free cells, the gap set by Warp's fixed broadphase overhead rather than the solver itself. Where FBF does spend its time is itself informative, since on the arch it is the inner cone solve, the honest cost of resolving a globally coupled contact problem, while on the house of cards it is an unoptimized host-side warm-start whose routine port to the GPU would recover most of the gap.

\section{Discussion and Limitations}
\label{sec:discussion}
For this solver, convergence of the outer iteration and physical correctness are two views of the same fact. On every benchmark, from a single sliding contact to the globally coupled arch, the outer iteration brings the dimensionless Coulomb residual of~\S\ref{sec:stopping} close to the tolerance on essentially every substep. On the house of cards, $93\%$ of substeps reach $r_c \le 10^{-6}$, and on the masonry arch the median residual is $1.1\times10^{-6}$, which is the single-precision floor identified in~\S\ref{sec:stopping} (Table~\ref{tab:convergence}). Because $r_c$ is a faithful surrogate for the Coulomb fixed-point deficit, this is the same result already shown in~\S\ref{sec:results}, where the structures settle into their physically correct configurations. A residual that does not quite reach $10^{-6}$ is then not a failure to resolve the contact problem. It is the iteration resting at the precision floor, where the physics is already right.

\begin{table}[t]
  \centering
  \caption{Convergence on the two contact-rich baselines ($\mu=0.8$), scored with the Coulomb residual $r_c$ of~\S\ref{sec:stopping}. Both scenes bring the bulk of substeps to the tolerance neighborhood.}
  \label{tab:convergence}
  \begin{tabular}{lcc}
    \hline
    scene & median $r_c$ & substeps with $r_c\!\le\!10^{-6}$ \\
    \hline
    house of cards & $6.3\times10^{-7}$ & $93\%$ \\
    masonry arch   & $1.1\times10^{-6}$ & $47\%$ \\
    \hline
  \end{tabular}
\end{table}

\begin{figure*}[t]
  \centering
  \includegraphics[width=\linewidth]{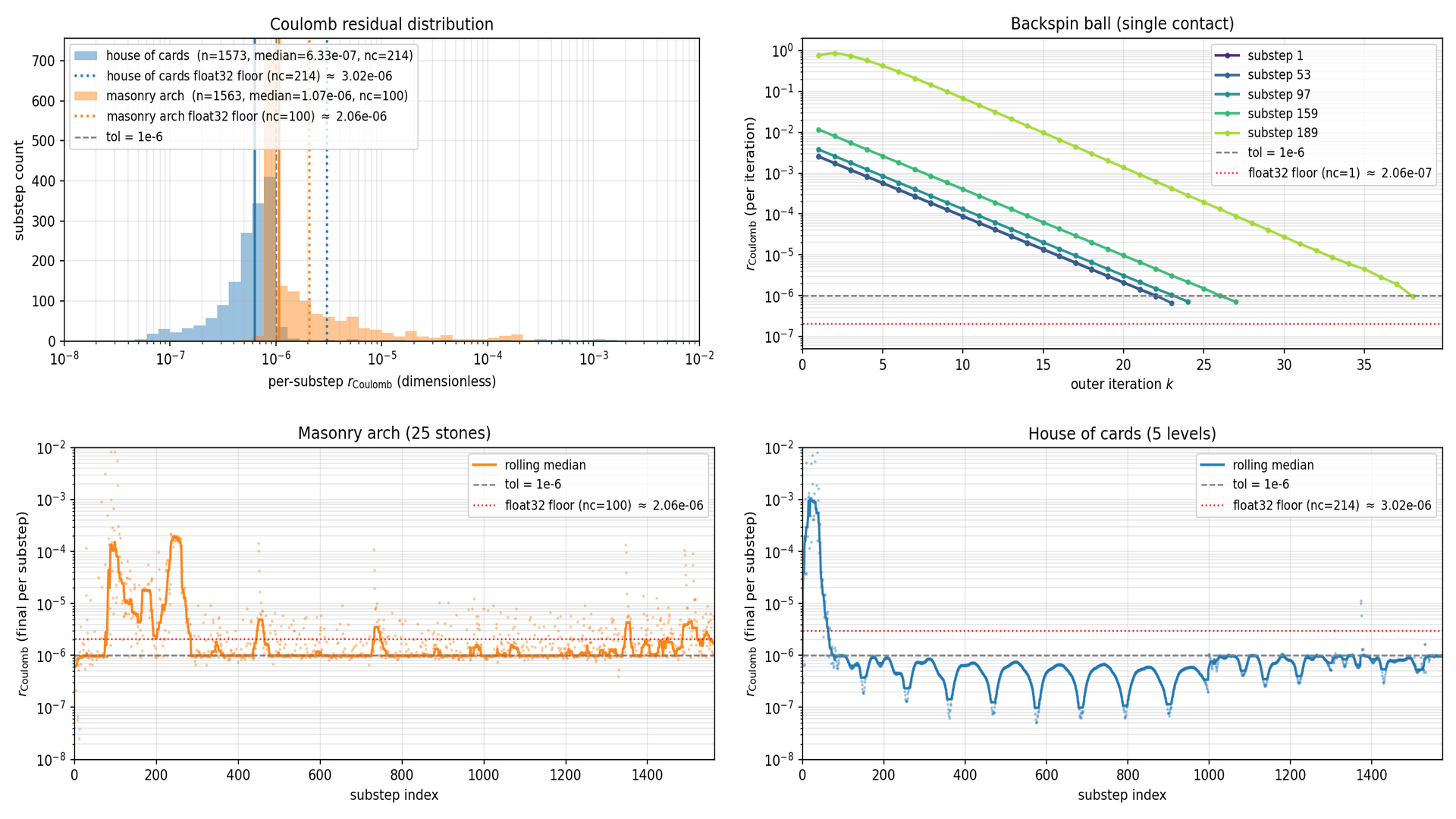}
  \caption{Convergence of the outer iteration measured by the Coulomb residual $r_c$~(\S\ref{sec:stopping}). The residual distribution (top left) places almost every substep of both contact-rich baselines at or below the single-precision floor. The single-contact backspin ball (top right) shows the geometric descent of $r_c$ with the outer iteration, reaching the tolerance within a few tens of iterations. The traces for the masonry arch (bottom left) and the house of cards (bottom right) follow the per-substep residual across the run, where it settles at the precision floor once each structure is loaded; the globally coupled arch is iteration-bound on its load-bearing substeps yet still reaches the floor.}
  \label{fig:convergence}
\end{figure*}

What governs this behavior is the one term the splitting treats explicitly, the non-associated De~Saxc\'{e} coupling $B$ of~\S\ref{sec:difficulty} and~\S\ref{sec:splitting}. When the influence of $B$ stays local, as for a single contact or the locally coupled house of cards, the explicit correction is contractive and the iteration converges in a handful of outer steps. When the contact graph instead carries force across the whole structure, as in the arch, $B$ couples the entire system and the explicit correction loses much of its contractivity, so the load-bearing substeps need many more iterations. Even on those substeps the residual still settles near the precision floor (Fig.~\ref{fig:convergence}).

Both control knobs behave as this picture predicts. Raising the outer-iteration budget tenfold helps only where the solver is iteration-bound. On the arch it brings more of the load-bearing substeps to the floor, while on the house of cards, which already converges in a few iterations, the extra budget does nothing, and in both cases the precision floor is the ceiling. The step size $\gamma$ sets how aggressively the explicit correction is applied. As $\gamma$ grows the residual falls monotonically up to a scene-dependent stability limit, and the adaptive rule of~\S\ref{sec:stepsize} reaches the best fixed $\gamma$ at lower cost. A small $\gamma$ is not a shortcut to convergence, because it stabilizes the iteration rather than the time step, and taken too small it fails to build up frictional support and the structure collapses under limited budgets. Finally, because $B$ is non-monotone, no fixed $\gamma$ recovers a formal convergence rate, since Tseng's guarantee does not transfer (\S\ref{sec:stepsize}). Convergence here is therefore established empirically, and the full budget and step-size sweeps are reported in Appendix~\ref{app:convergence}.

The part of the residual that remains is the dual feasibility term $\varepsilon_{\mathrm{vel}}$ of~\S\ref{sec:stopping}, the velocity side on which the De~Saxc\'{e} coupling acts. This is where the global regime pays for its weaker contraction, with the arch settling a little above the house of cards, though both stay below the single-precision wall, and separating the two would require a float64 pipeline. Two checks confirm that this residual is a property of the outer iteration and not an implementation artifact. Replacing the inner block Gauss--Seidel solver with Clarabel changes the cost but not the outer residual, so the inner cone QP is not the bottleneck. The safeguarded step-size rule stays largely inactive on the hard runs, so the residual is not a step-size instability (Appendix~\ref{app:convergence}). What remains open is a formal convergence rate for the outer iteration under the non-monotone coupling.

We develop the architecture for rigid bodies and commit to the contact law at the velocity level, and deformable simulation lies outside this scope for two related reasons. The velocity-level law resolves the contact impulses and the resulting velocities rather than an interpenetration-free trajectory across the step, which deformable solvers secure at the position level through barrier potentials or continuous collision detection. The reduced contact-space solve is also efficient only because a rigid body's mass inverts independently per body, whereas a deformable body couples that inverse into a global elastic solve. Extending the splitting to deformable contact would therefore pair it with a position-level treatment, which we leave to future work.

\revoff
\section{Conclusion}
\label{sec:conclusion}
We have presented a splitting architecture for the exact reduced Coulomb friction law. Starting from the cone-QP / scalar-coupling decomposition of Acary~et~al.~\cite{acary2011formulation}, we replace their Picard outer iteration with a Tseng-style forward--backward--forward scheme and pair it with a safeguarded adaptive step size tuned to the non-monotone De~Saxc\'{e} and Feng coupling. The resulting structure makes the inner cone solve modular: the outer loop commits to the exact law, while any strongly convex SOCP solver can be substituted inside. A matrix-free, contact-parallel implementation on top of Warp and Newton demonstrates the architecture on rigid-body benchmarks. On single-contact and low-coupling problems, the method reproduces the exact law to near-analytical accuracy, where the convex-relaxed CCP solver MuJoCo exhibits the characteristic normal-separation and creep artifacts. On contact-rich structures (house of cards, masonry arch) the method preserves local frictional support and global arch equilibrium through projectile impact. 

The main limitation of the current architecture, diagnosed in~\S\ref{sec:discussion}, is that the explicit outer treatment of the non-associated coupling loses effective contractivity on globally coupled contact graphs, so that complementarity is not fully resolved within the outer budget. Addressing this regime with the insights of the analysis in~\S\ref{sec:discussion} through a partial implicit treatment of the coupling is a natural next step. Our modular architecture is designed to accommodate such experiments, for example, adaptive mode detection or accelerated outer iteration, without disturbing the inner cone solver or the overall splitting structure.
Despite this limitation, and that our solver is un-optimized academic software, our method appears to perform better in contact-rich scenarios (e.g., tall arch with 101 blocks) than the recently released (March/April 2026) industrial-strength software Kamino~\cite{kamino2026}, and similarly in simpler scenarios. Open-source code will be available at \url{http://www.cs.ubc.ca/research/fbf-friction}.

\section*{Acknowledgments}
This work was supported, in part, by NSERC Discovery Grants to Ascher and Pai.
The masonry-arch geometry is from the Rigid-IPC dataset~\cite{ferguson2021intersection} (\url{https://github.com/ipc-sim/rigid-ipc}).

\bibliographystyle{eg-alpha-doi}
\bibliography{refs}

\appendix
\revon
\section{Convergence Analysis}
\label{app:convergence}

This appendix expands the convergence summary of~\S\ref{sec:discussion} on the two contact-rich baselines, the house of cards and the masonry arch (both at $\mu=0.8$), scored with the Coulomb residual $r_c$ of~\S\ref{sec:stopping}.

\paragraph*{Outer-budget sensitivity.}
Raising the outer-iteration cap from $200$ to $2000$ separates the iteration-bound regime from the converged one (Table~\ref{tab:budget}). The arch saturates the $200$ cap on $51\%$ of substeps; the tenfold budget lifts the share at $r_c\le10^{-6}$ from $47\%$ to $61\%$ and suppresses the transient excursion on the load-bearing substeps, at a proportional cost in wall time. The house of cards already terminates at a median of five outer iterations, so the extra budget is unused. In both cases the median residual remains bounded below by the float32 floor.

\begin{table}[t]
  \centering
  \caption{Outer-budget sensitivity: outer-iteration cap $200$ (\texttt{std}) versus $2000$ (\texttt{10x}). The residual, share, and iteration count are medians over substeps; timing is wall-clock per frame.}
  \label{tab:budget}
  \small
  \begin{tabular}{l c c c r}
    \hline
    run & median $r_c$ & $r_c\!\le\!10^{-6}$ & iters & ms/frame \\
    \hline
    cards \texttt{std} & $6.3\times10^{-7}$ & $93\%$ & 5    & 199 \\
    cards \texttt{10x} & $1.5\times10^{-5}$ & $3\%$  & 40   & 449 \\
    arch \texttt{std}  & $1.1\times10^{-6}$ & $47\%$ & 200  & 595 \\
    arch \texttt{10x}  & $1.0\times10^{-6}$ & $61\%$ & 1304 & 4539 \\
    \hline
  \end{tabular}
\end{table}

\begin{figure*}[t]
  \centering
  \includegraphics[width=\linewidth]{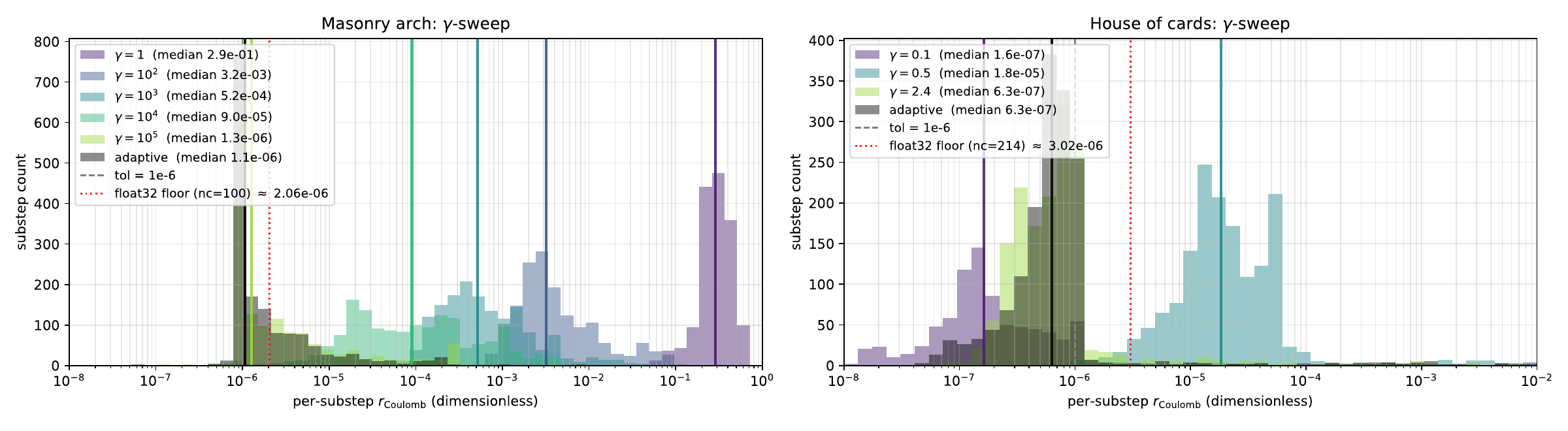}
  \caption{Step-size sweep on the two contact-rich baselines, scored by the per-substep Coulomb residual $r_c$ of~\S\ref{sec:stopping}. On the masonry arch (left) the distribution marches monotonically toward the single-precision floor as $\gamma$ grows, with $\gamma=10^5$ and the adaptive rule of~\S\ref{sec:stepsize} both resting at the floor. On the house of cards (right) the adaptive rule and $\gamma=2.4$ settle at the floor while $\gamma=0.5$ converges more slowly, and the smallest $\gamma=0.1$ reaches an even lower residual yet collapses the stack (Table~\ref{tab:gamma}), which shows that a small $\gamma$ stabilizes the iteration rather than the time step. The two largest card step sizes diverge (Table~\ref{tab:gamma}) and are omitted. Vertical markers give each run's median, the $10^{-6}$ tolerance, and the float32 floor $\epsilon_{32}\sqrt{3n_c}$.}
  \label{fig:gamma}
\end{figure*}

\paragraph*{Step-size sweep.}
Sweeping a fixed $\gamma$ over several orders of magnitude (Table~\ref{tab:gamma}, Figure~\ref{fig:gamma}) shows $r_c$ decreasing monotonically with $\gamma$ on the arch, with $\gamma=10^5$ essentially reproducing the adaptive baseline, both at the floor, while no fixed $\gamma$ closes the formal gap. On the house of cards the smallest $\gamma$ attains the lowest residual yet collapses the stack (a $9.7\,\mathrm{m}$ drop), which confirms that a small $\gamma$ stabilizes the iteration but not the time step, and the largest values diverge. The adaptive rule of~\S\ref{sec:stepsize} matches the best fixed $\gamma$ on both scenes at lower iteration cost.

\begin{table}[t]
  \centering
  \caption{Fixed-$\gamma$ sweep on each scene, with the adaptive rule of~\S\ref{sec:stepsize} in the final row. For the house of cards, \texttt{max drop} is the largest card displacement, a settling-failure indicator, and NaN marks a diverged run.}
  \label{tab:gamma}
  \begin{tabular}{c c | c c c}
    \hline
    \multicolumn{2}{c|}{masonry arch} & \multicolumn{3}{c}{house of cards} \\
    $\gamma$ & median $r_c$ & $\gamma$ & median $r_c$ & max drop \\
    \hline
    $1$      & $2.9\times10^{-1}$ & $0.1$ & $1.6\times10^{-7}$ & $9.7\,\mathrm{m}$ \\
    $10^2$   & $3.2\times10^{-3}$ & $0.5$ & $1.8\times10^{-5}$ & $0.4\,\mathrm{m}$ \\
    $10^3$   & $5.2\times10^{-4}$ & $2.4$ & $6.3\times10^{-7}$ & $0.3\,\mathrm{m}$ \\
    $10^4$   & $9.0\times10^{-5}$ & $7.5$ & NaN & --- \\
    $10^5$   & $1.3\times10^{-6}$ & $30$  & NaN & --- \\
    \hline
    adaptive & $1.1\times10^{-6}$ & adaptive & $6.3\times10^{-7}$ & $0.3\,\mathrm{m}$ \\
    \hline
  \end{tabular}
\end{table}

\paragraph*{Float32 precision floor.}
The pipeline is single precision throughout, so the dual-feasibility and complementarity components of $r_c$ carry a rounding floor of order $\epsilon_{32}\sqrt{3 n_c}$, with $\epsilon_{32}\approx1.2\times10^{-7}$ the single-precision unit roundoff and $n_c$ the active-contact count; at $n_c\approx108$ this is $\approx2.1\times10^{-6}$. On both baselines the median $r_c$ sits within a factor of two of this estimate, so $\varepsilon=10^{-6}$ is the tightest tolerance meaningful in float32, and a tighter threshold would test rounding noise rather than convergence.

\paragraph*{Diagnostics.}
Two controls confirm the remaining residual is a property of the outer iteration. Replacing the inner block Gauss--Seidel solver with the interior-point solver Clarabel raises per-subproblem accuracy at substantially higher cost but leaves the outer residual unchanged, so the inner cone QP is not the bottleneck. And on the hard arch runs the safeguarded step-size rule is largely inactive, with little to no $\gamma$ adaptation, so the residual is not a step-size instability.

\section{Performance}
\label{app:performance}

This appendix reports the per-benchmark timing and iteration data (Table~\ref{tab:perf-full}) behind the cost summary of~\S\ref{sec:performance}. All three solvers run at a matched residual tolerance $\varepsilon=10^{-6}$: FBF on the dimensionless Coulomb residual $r_c$, MuJoCo on its native tolerance, and Kamino on the PADMM primal, dual, and complementarity residuals. FBF uses the adaptive step size of~\S\ref{sec:stepsize} with a $200$-iteration outer cap and a fixed budget of $10$ inner block Gauss--Seidel sweeps per outer step, raised to $30$ on the arch; MuJoCo runs its Newton solver with an elliptic cone at $500$ iterations; Kamino uses a blocked Cholesky factorization under a $200$-iteration PADMM cap, switching to its matrix-free conjugate-residual path on the house of cards, where the blocked factor over-sizes the contact graph. The CPU runs (Tables~\ref{tab:perf-full} and~\ref{tab:perf-large}) are single precision on a single Apple-silicon Mac with the Warp and Newton backend, executed on the CPU and run sequentially so the timings are uncontended; the GPU runs (Table~\ref{tab:perf-gpu}) use the same backend on an NVIDIA GeForce RTX~3090. In both cases the reported time is the mean wall time per step over the run; for Kamino on the house of cards, whose matrix-free path pays a large one-time first-step kernel compile and CUDA-graph capture, that first step is excluded as warmup.

\begin{table}[t]
  \centering
  \caption{Per-benchmark wall time and iteration counts at matched tolerance $\varepsilon=10^{-6}$. FBF time is the mean wall time per step; the MuJoCo and Kamino columns are relative to FBF. \texttt{outer} is the median and $95$th-percentile outer-iteration count per substep, and \texttt{inner} the fixed block Gauss--Seidel sweeps per outer step. On the house of cards Kamino's blocked factor over-sizes the contact graph, so the listed factor is its slower matrix-free path, a lower bound measured before the run was stopped. The super-critical turntable cells eject the cube, leaving contact-free substeps ($n_c=0$).}
  \label{tab:perf-full}
  \footnotesize
  \setlength{\tabcolsep}{3pt}
  \begin{tabular}{l c r c c c c}
    \hline
    benchmark & $n_c$ & FBF (ms) & MuJoCo & Kamino & outer & inner \\
    \hline
    backspin                  & $1$   & $6.0$ & $0.3\times$  & $1.8\times$  & $25/30$   & $10$ \\
    incline $\mu0.5$          & $4$   & $5.5$ & $0.4\times$  & $3.1\times$  & $5/5$     & $10$ \\
    incline $\mu0.4$          & $4$   & $5.3$ & $0.6\times$  & $4.1\times$  & $5/5$     & $10$ \\
    painlev\'{e} $\mu0.5$     & $4$   & $7.0$ & $0.9\times$  & $1.1\times$  & $5/25$    & $10$ \\
    painlev\'{e} $\mu0.55$    & $4$   & $6.4$ & $0.6\times$  & $0.7\times$  & $5/30$    & $10$ \\
    turntable $\mu0.5\omega2$ & $4$   & $6.8$ & $0.6\times$  & $0.6\times$  & $10/15$   & $10$ \\
    turntable $\mu0.5\omega5$ & $0$   & $3.1$ & $1.2\times$  & $1.1\times$  & $0/15$    & $10$ \\
    turntable $\mu0.2\omega2$ & $0$   & $3.7$ & $1.1\times$  & $0.9\times$  & $0/15$    & $10$ \\
    turntable $\mu0.2\omega5$ & $0$   & $3.1$ & $1.2\times$  & $1.5\times$  & $0/10$    & $10$ \\
    house of cards            & $214$ & $199$ & $8.7\times$  & $\ge\!50\times$ & $5/115$   & $10$ \\
    masonry arch              & $100$ & $595$ & $23.3\times$ & $0.38\times$ & $200/200$ & $30$ \\
    \hline
  \end{tabular}
\end{table}

Eight of the eleven cells run at a median of at most ten outer iterations, and only the globally coupled arch saturates the $200$-iteration cap, the same iteration-bound regime identified in Appendix~\ref{app:convergence}. The backspin ball, reported in the submission at sixty or more outer iterations, converges at a median of twenty-five once the step size is tuned and the dimensionless residual of~\S\ref{sec:stopping} is used as the stopping test, in line with the other few-contact scenes. The house of cards and the arch bracket the cost. On the arch the bulk of each substep is the inner cone solve, the honest price of resolving a globally coupled contact problem, whereas on the house of cards the dominant term is an unoptimized host-side warm-start whose median cost grows linearly with the contact count and which a routine port to a Warp kernel would largely remove.

\paragraph*{Large scale.}
The two large-scale scenes of~\S\ref{sec:performance}, the $101$-stone arch and the ten-level house of cards, separate the solvers by completion rather than by margin (Table~\ref{tab:perf-large}). FBF completes both, while MuJoCo finishes neither within a wall-clock budget many times FBF's, and Kamino completes only the ten-level house of cards, and only through its slower matrix-free path, while integrating a physically wrong, contact-free trajectory on the arch.

\begin{table}[t]
  \centering
  \caption{Large-scale results at $\varepsilon=10^{-6}$, $\mu=0.8$. FBF time is the mean wall time per step. MuJoCo does not finish (DNF) either scene; the bracketed factor is a lower bound on its slowdown from the pace it reached before being stopped. Kamino runs the house of cards only through its slower matrix-free path, the listed factor a lower bound of the same kind, and produces a contact-free trajectory on the arch.}
  \label{tab:perf-large}
  \small
  \setlength{\tabcolsep}{4.5pt}
  \begin{tabular}{l r l l}
    \hline
    scene & FBF (ms) & MuJoCo & Kamino \\
    \hline
    arch, $101$ stones          & $1234$ & DNF ($\ge\!30\times$) & contact-free \\
    house of cards, $10$ levels & $853$  & DNF ($\ge\!68\times$) & $\ge\!70\times$ \\
    \hline
  \end{tabular}
\end{table}

\paragraph*{GPU.}
Table~\ref{tab:perf-gpu} repeats the four contact-rich scenes on the GPU, under the same parameters and matched tolerance, reporting the mean wall time per step and the physical outcome of each solver. On the GPU the comparison turns on physical correctness rather than completion, as MuJoCo and Kamino now advance the large scenes they could not finish on the CPU (Table~\ref{tab:perf-large}). FBF is the only solver that preserves every structure. MuJoCo is the fastest throughout but topples or collapses all four, while Kamino now advances all four as well, its matrix-free path keeping the two houses of cards standing, though it still collapses the $101$-stone arch. FBF is the slowest on the arches, where its inner cone solve is a sequence of small per-color Gauss--Seidel batches with a per-contact projection that underutilizes the GPU relative to the dense batched updates MuJoCo and Kamino are built around; on the dense card stacks Kamino's matrix-free path is instead the slowest, several times slower than FBF. These GPU timings are an unoptimized port: the host-side warm-start and the serial per-color inner sweep are unchanged from the CPU build, so FBF runs slower on the GPU than on a single CPU core (Tables~\ref{tab:perf-full} and~\ref{tab:perf-large}). Kernelizing the warm-start, as noted above, together with a batched inner solve would be needed to make the GPU port competitive; we report these timings for completeness.

\begin{table}[t]
  \centering
  \caption{GPU performance on the four contact-rich scenes, run with the same parameters and matched tolerance ($\varepsilon=10^{-6}$, $\mu=0.8$) as the CPU experiments (Table~\ref{tab:perf-large}). Each cell gives the mean wall time per step in milliseconds on an NVIDIA GeForce RTX~3090 and the physical outcome: \emph{stands} preserves the structure, while \emph{topples} and \emph{collapses} fail it physically. FBF is the only solver that preserves every structure; MuJoCo is fastest but loses all four; Kamino keeps the houses of cards standing through its matrix-free path but collapses the $101$-stone arch. The Kamino card timings exclude its one-time first-step kernel compile and CUDA-graph capture ($1.5\,\mathrm{s}$ and $353\,\mathrm{s}$ at $5$ and $10$ levels).}
  \label{tab:perf-gpu}
  \small
  \setlength{\tabcolsep}{6pt}
  \begin{tabular}{l c c c}
    \hline
    scene & FBF (ms) & MuJoCo (ms) & Kamino (ms) \\
    \hline
    card, $5$ levels   & $479$, stands    & $37$, topples    & $4126$, stands \\
    card, $10$ levels  & $1513$, stands   & $621$, topples   & $4285$, stands \\
    arch, $25$ stones  & $2051$, stands   & $61$, collapses  & $355$, stands \\
    arch, $101$ stones & $3731$, stands   & $821$, collapses & $615$, collapses \\
    \hline
  \end{tabular}
\end{table}
\revoff
\end{document}